\documentclass[11pt,a4paper]{article}
\usepackage[noadjust]{cite}

\usepackage[margin=2.8cm,bottom=3.5cm]{geometry}

\usepackage{amsmath, amssymb}
\usepackage{color}
\pdfoutput=1 

\usepackage{graphicx} 
\usepackage{subcaption}
\usepackage[T1]{fontenc} 

\title{Kink-antikink collisions in the $\phi^8$ model: short-range to long-range journey}
\author{Dionisio Bazeia \\
        Departamento de F\'isica, Universidade Federal da Para\'iba,\\
        Jo\~ao Pessoa - PB - 58051-970, Brazil\\
        bazeia@fisica.ufpb.br
            \and
        Jo\~ao G. F. Campos \\
        Departamento de F\'isica, Universidade Federal da Para\'iba,\\
        Jo\~ao Pessoa - PB - 58051-970, Brazil\\
        joaogfc@gmail.com
            \and
        Azadeh Mohammadi\\
        Departamento de F\'isica, Universidade Federal de Pernambuco,\\
        Av. Prof. Moraes Rego, 1235, Recife - PE - 50670-901, Brazil\\
        azadeh.mohammadi@ufpe.br}

\begin{document}

\maketitle
\begin{abstract}
We studied kink-antikink collisions in (1+1)-dimensional spacetime for all $Z_2$ symmetric $\phi^8$ models with four degenerate minima. Such a polynomial model has only one free parameter, allowing us to conduct an exhaustive analysis. We performed detailed simulations in all three sectors of the model. We observed resonance windows from both localized and delocalized modes, as well as a sector change with the formation of additional kink-antikink pairs.
Furthermore, we were able to show how collisions are modified when two quadratic minima merge into a quartic one, causing the kinks to acquire a long-range character. We demonstrated that when the tail not facing the opposing kink is long-range, incoming kinks and antikinks decay directly into radiation, as suggested in \cite{campos2021interaction}, by forming a large number of small kink-antikink pairs. Finally, we briefly discussed whether our analysis could be generalized to other polynomial models.

\end{abstract}

\section{Introduction}

Physics seeks to understand wave equations, which describe phenomena such as waves in fluids, light propagation, and strings under tension. With access to the quantum realm, wave equations have also been found to describe particles and fields. In particular, some nonlinear wave equations have fascinating structures which are localized and propagate without losing their shape \cite{rajaraman1982solitons, manton2004topological}. These can be topological or non-topological and sometimes are called solitary waves or solitons.

In (1+1)-dimensional spacetime, localized structures known as kinks exist, which connect different scalar field vacua. They appear in many systems that are effectively unidimensional in space and have degenerate minimum energy configurations. Examples include electron displacement in polyacetylene \cite{su1979solitons, bernasconi2015chaotic}, DNA properties \cite{yakushevich2006nonlinear}, Josephson junctions \cite{ustinov1998solitons}, domain walls in ferromamgnets \cite{kardar2007statistical}, gold dislocations \cite{el1987double}, properties of Rydberg atoms \cite{saffman2010quantum}, graphene deformations \cite{yamaletdinov2017kinks, martin2021fractal}, and branes \cite{rubakov1983we, khoury2001ekpyrotic}.

The sine-Gordon model is the prototype of an integrable model that features elastic kink collisions, while the $\phi^4$ model is the prototype of a non-integrable model that exhibits inelastic collisions. Interestingly, the $\phi^4$ model exhibits a resonance phenomenon, as is now well known after a series of pioneering works \cite{sugiyama1979kink, campbell1983resonance, peyrard1983kink, campbell1986kink, anninos1991fractal}. It consists of a kink and an antikink temporarily trapped after an initial bounce and separating in subsequent bounces. In short, it occurs because there is an exchange between the kink's translational and vibrational degrees of freedom. This picture is now better understood in terms of a collective coordinate model \cite{manton2021kink, manton2021collective}.

Recently, resonance and other exciting phenomena in kink-antikink collisions have been investigated by many authors. Examples consist of polynomial models \cite{dorey2011kink, gani2014kink, gani2015kink, izquierdo2021scattering}, hyperbolic models \cite{bazeia2018scattering, bazeia2019kink, bazeia2020oscillons, marjaneh2022collisions}, multikink collisions \cite{marjaneh2017high, marjaneh2017multi, marjaneh2018extreme, gani2021exotic}, multifield collisions \cite{halavanau2012resonance, alonso2018reflection, alonso2020non, alonso2021kink}, the double sine-Gordon model \cite{gani2018scattering, gani2019multi, simas2020solitary, campos2021wobbling}, long-range \cite{manton2019forces, christov2019kink,christov2019long,christov2021kink, campos2021interaction} and compact \cite{bazeia2019scattering, bazeia2021semi} kinks, and interactions with boundaries \cite{arthur2016breaking,dorey2017boundary,lima2019boundary}. Interestingly, the exchange mechanism can occur via the interaction of other types of oscillating modes besides the kink's shape mode. These can be delocalized modes \cite{dorey2011kink, christov2021kink, bazeia2021semi}, sphalerons \cite{adam2021sphalerons}, quasinormal modes \cite{dorey2018resonant, campos2020quasinormal} and fermion bound states \cite{bazeia2022resonance}. In particular, there have been great advances in the collective coordinate description of the resonance mechanism of the $\phi^6$ model via delocalized modes \cite{adam2022multikink}.

A unique property of scalar fields in (1+1)-dimensional spacetime is their dimensionless nature, which allows polynomials of any order in the scalar field potential when constructing the Lagrangian. We are interested in potentials with $\phi\to-\phi$ symmetry\footnote{See Refs.~\cite{amado2020phi, lima2021scattering} as examples of how this property can be relaxed.}. The $\phi^4$ model with symmetry breaking has no free parameters after rescaling and has been studied extensively. The $\phi^6$, on the other hand, has a single free parameter and is typically analyzed at the first-order phase transition. At that point, it has also been extensively studied. From an experimental point of view, some parameters need to be tuned to maintain the system there. Thus, one also needs to study the system's behavior away from the first-order phase transition. Luckily, it can be inferred by the study of a model proposed in Ref.~\cite{dorey2021resonance}, which also exhibits a first-order phase transition. There, the authors studied the possibility of moving away from it. Therefore, it is possible to say that the resonance mechanism of the $\phi^4$ and $\phi^6$ models are generally better understood compared with the higher-order polynomial models.

One may wonder what new phenomena will arise in kink-antikink collisions as we increase the order of the polynomial and whether it is worthwhile to do so. Specifically, we aim to investigate the $\phi^8$ model to understand what new phenomena emerge. Kink-antikink interactions in $\phi^8$ models with long-range tails reveal highly nontrivial new phenomena \cite{christov2019kink, christov2019long, christov2021kink, campos2021interaction}. See also Refs.~\cite{bazeia2018analytical, gani2020explicit, blinov2022kinks} for works where kinks with long-range tails are taken into account and Ref. \cite{khare2022kink} for a recent review. However, we focus here on how the system approaches the long-range regime and other phenomena unrelated to the long-range property. While a few other works, such as \cite{gani2015kink, gani2021exotic}, have explored kink-antikink collisions in the $\phi^8$ model, an exhaustive analysis of the possible behaviors is still lacking in the literature. This gap is what we aim to address in the present work.

Polynomial field potentials of orders 8, 10, and 12, along with their kink solutions, have been listed in \cite{khare2014successive}, while quantum corrections to $\phi^8$ kinks were computed in \cite{takyi2020quantum}. In Ref.~\cite{khare2014successive}, the authors show that the $\phi^8$ potential has two free parameters. This means that an exhaustive analysis is impractical. However, there is only one free parameter at the first-order phase transition, with four degenerate minima, making the analysis doable. We will pursue this goal in the present work. To obtain a $\phi^8$ model with four degenerate minima, a single parameter has to be tuned. Therefore, at least in theory, it should not be more difficult to deal with such a system at the first-order phase transition than a $\phi^6$ system in general. In section~\ref{sec:model}, we discuss some properties of the $\phi^8$ model with four minima. In section~\ref{sec:Results}, we show the results of several types of kink-antikink collisions, and we conclude our work in section~\ref{sec:conclusion}.
 
\section{Model}
\label{sec:model}

We aim to study the following scalar field model
\begin{equation}
    \mathcal{L}=\frac{1}{2}\partial_\mu\phi\partial^\mu\phi-V_b(\phi).
\end{equation}
The current work focuses on the $\phi^8$ model with four degenerate minima. After rescaling, they can be described by the following family of potentials with one free parameter
\begin{equation}\label{pot}
    V_b(\phi)=\frac{1}{2}(\phi^2-1)^2(\phi^2-b)^2.
\end{equation}
We will consider the case where $b>0$. Then, the minima are at $\phi=\pm 1$ and $\phi=\pm\sqrt{b}$. Moreover, the regions $0<b<1$ and $b>1$ are physically equivalent by scaling arguments, so we will take $b\in(0,1)$. 

The equations of motion are given by
\begin{equation}
    \phi_{tt}-\phi_{xx}+V_b^\prime(\phi)=0.
\end{equation}
The Bogomol'nyi–Prasad–Sommerfield (BPS) equation can be used to obtain the kink and antikink solutions
\begin{equation}
    \phi_x=\pm(\phi^2-1)(\phi^2-b).
    \label{eq:BPS}
\end{equation}
Implicit analytical solutions to the BPS equation can be obtained as shown in Ref.~\cite{khare2014successive}. For our purposes, it is enough to compute the kink and antikink solutions numerically, which is a simple task.

\begin{figure}
    \centering
    \begin{subfigure}{0.85\textwidth} 
    \includegraphics[width=\textwidth]{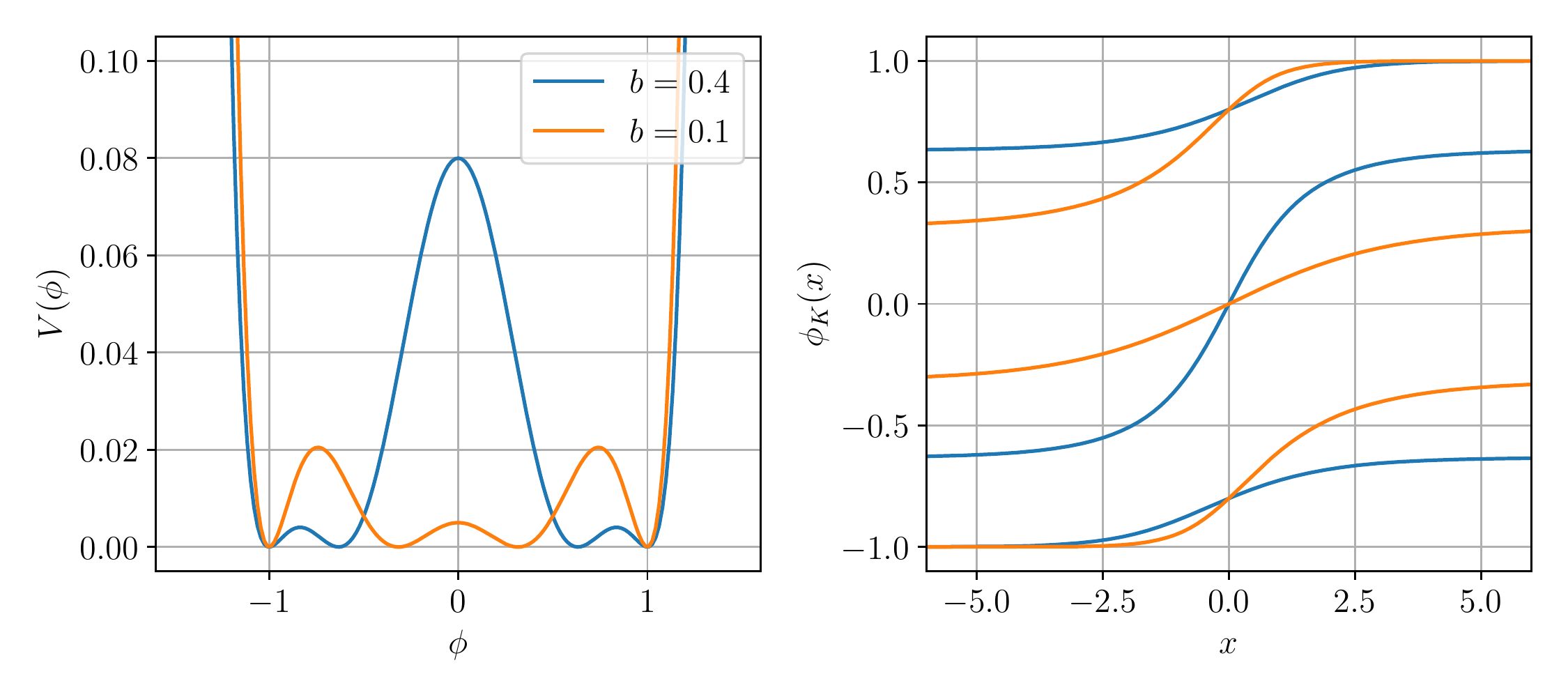}
    \end{subfigure}
    \begin{subfigure}{0.85\textwidth} 
    \includegraphics[width=\textwidth]{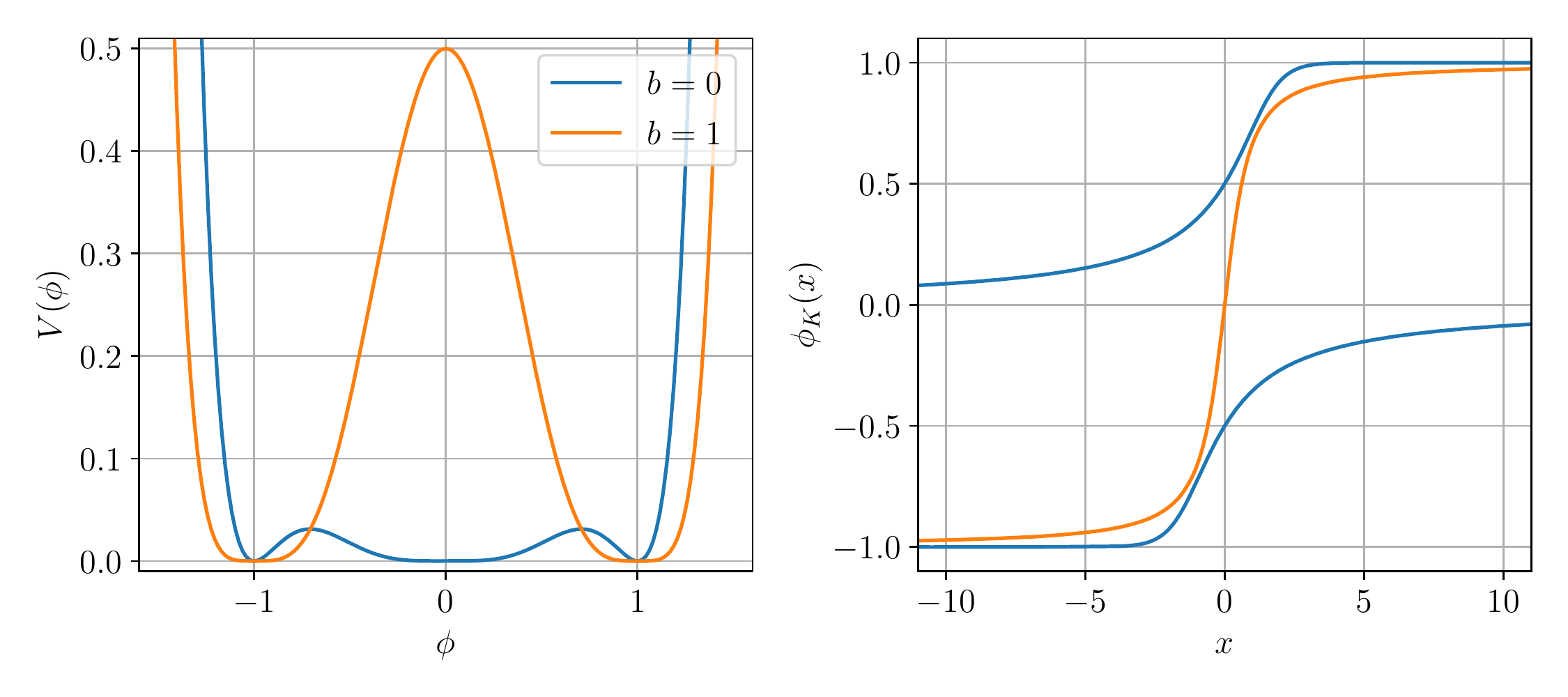}
    \end{subfigure}
    \caption{Potential and kink solutions for several values of $b$.}
    \label{fig:pot1}
\end{figure}


In the top row of Fig.~\ref{fig:pot1}, we show two potentials with $b$ in the range $0<b<1$ and their corresponding kinks. They possess two inner minima at $\phi=\pm\sqrt{b}$ and two outer minima at $\phi=\pm 1$. Thus, there are two different sectors. The inner sector corresponds to solutions that interpolate between the two inner minima. It contains a symmetric inner kink and its corresponding antikink. The outer sectors correspond to solutions that interpolate between an inner minimum and an outer minimum. They contain two asymmetric outer kinks and their corresponding antikinks. The kinks' profiles are denoted by $\phi_K$. To specify the sector, we use a superscript $(i)$ for the inner one and $(o)$ for the outer one. For antikinks, we put a bar over the subscript $K$.

The cases with $b=7-4\sqrt{3}$ and $b=1/4$ were studied in Refs. \cite{gani2015kink} and \cite{gani2021exotic}, respectively.
In the bottom row of figure \ref{fig:pot1}, we show the two limiting cases where $b=0$ and $b=1$. Note that some of the minima are much flatter than the ones in the previous graphs, leading to kinks with long-range character. There are two asymmetric kinks and their corresponding antikinks for $b=0$. They all possess a short-range tail on one side and a long-range tail on the other side. This case was studied in detail in \cite{christov2019kink, christov2019long, christov2021kink}. The authors showed that the model exhibits resonance windows when the long-range tails face each other. The exchange mechanism occurs because the kink-antikink pair possesses vibrational modes. For $b=1$, there is a single symmetric kink with a double long-range tail and its corresponding antikink. This case was studied in detail in \cite{campos2021interaction}. The authors showed that there are no resonance windows in kink-antikink collisions and that the critical velocity is ultra-relativistic. The explanation for this phenomenon comes from the long-range character of the tail that is not facing the opposing kink, as we will show shortly.
From the discussion above, we see that by varying the parameter $b$, we can explore the scattering behavior of kinks that gradually acquire a long-range character. Therefore, studying such systems may shed more light on the scattering behavior of long-range kinks in general.

\begin{figure}
    \centering
    \includegraphics[width=0.8\textwidth]{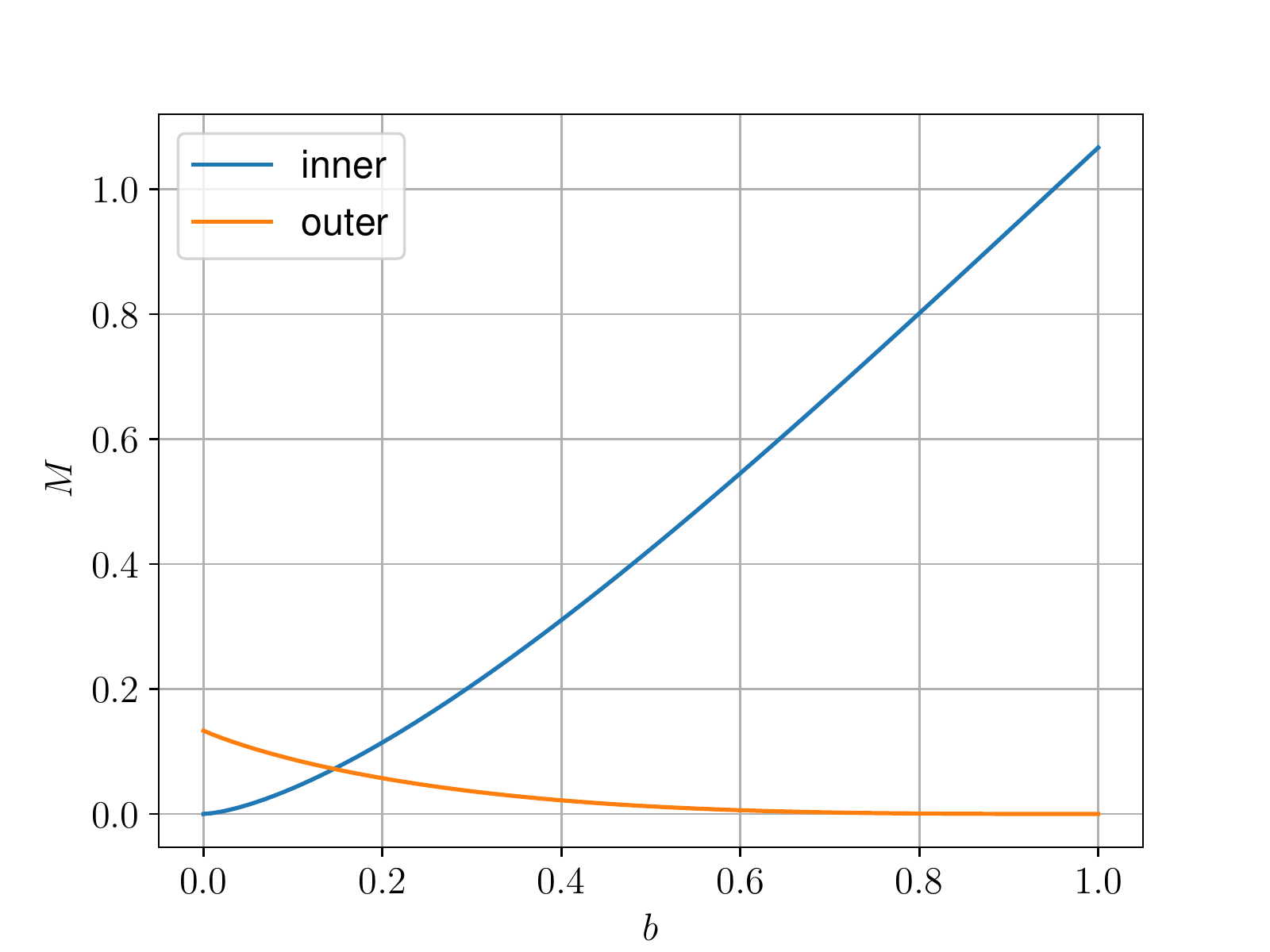}
    \caption{Masses of inner and outer kinks as a function of $b$.}
    \label{fig:mass}
\end{figure}

Although the kinks will be computed numerically, their masses can be computed analytically. The potential in \eqref{pot} can be written as
\begin{equation}
V_b(\phi)=\frac12 \left(\frac{dW_b}{d\phi}\right)^2,
\end{equation}
where $W_b=W_b(\phi)$ is a function of the scalar field with the form
\begin{equation}
    W_b(\phi)=b\,\phi -\frac13 (1+b)\,\phi^3+\frac15 \phi^5.
\end{equation}
For static configurations, the corresponding energy density $\rho_b(x)$ can be written in the following form
\begin{equation}
    \rho_b(x)=\frac12\phi_x^2+\frac12\left(\frac{dW_b}{d\phi}\right)^2=\pm\frac{dW_b}{dx}+\frac12\left(\phi_x\mp \frac{dW_b}{d\phi}\right)^2.
\end{equation}
Since the kink solutions obey the first-order equations in \eqref{eq:BPS}, their masses are given by the energies of the static solutions at the respective sectors, i.e., $M_b=|W_b(\phi(\infty))-W_b(\phi(-\infty))|$. Thus, they can be computed as $M_b^{(i)}=|W_b(\sqrt{b})-W_b(-\sqrt{b})|$ and $M_b^{(o)}=|W_b(1)-W_b(\sqrt{b})|$ resulting in
\begin{equation}
    M_b^{(i)}=\frac{4}{15}b^{3/2}(5-b),\quad M_b^{(o)}=\frac{2}{15}(1-\sqrt{b})^3(1+3\sqrt{b}+b),
\end{equation}
reproducing the results obtained in Ref.~\cite{khare2014successive}.
Fig.~\ref{fig:mass} represents them graphically.

The scattering behavior of kink-antikink systems depends on the presence of vibrational modes. Therefore, studying the kink's stability equation, which describes such modes, is essential.
Substituting the ansatz $\phi(x,t)=\phi_K(x,t)+\eta(x)\cos(\omega t)$ in the equation of motion, we obtain a Schr\"{o}dinger-like equation
\begin{equation}
    -\eta^{\prime\prime}(x)+U(x)\eta(x)=\omega^2\eta(x).
\end{equation}
The effective potential is defined as $U(x)\equiv V_b^{\prime\prime}(\phi_K(x))$. It is well known that this equation has no negative eigenvalues, and thus, the kink is stable. Due to the translational invariance of the model, the kink possesses a zero mode which is proportional to the derivative of the solution itself, given by
\begin{equation}
    \eta_0(x)\propto\phi_K^\prime(x).
\end{equation}


\begin{figure}
    \centering
    \begin{subfigure}{0.85\textwidth} 
    \includegraphics[width=\textwidth]{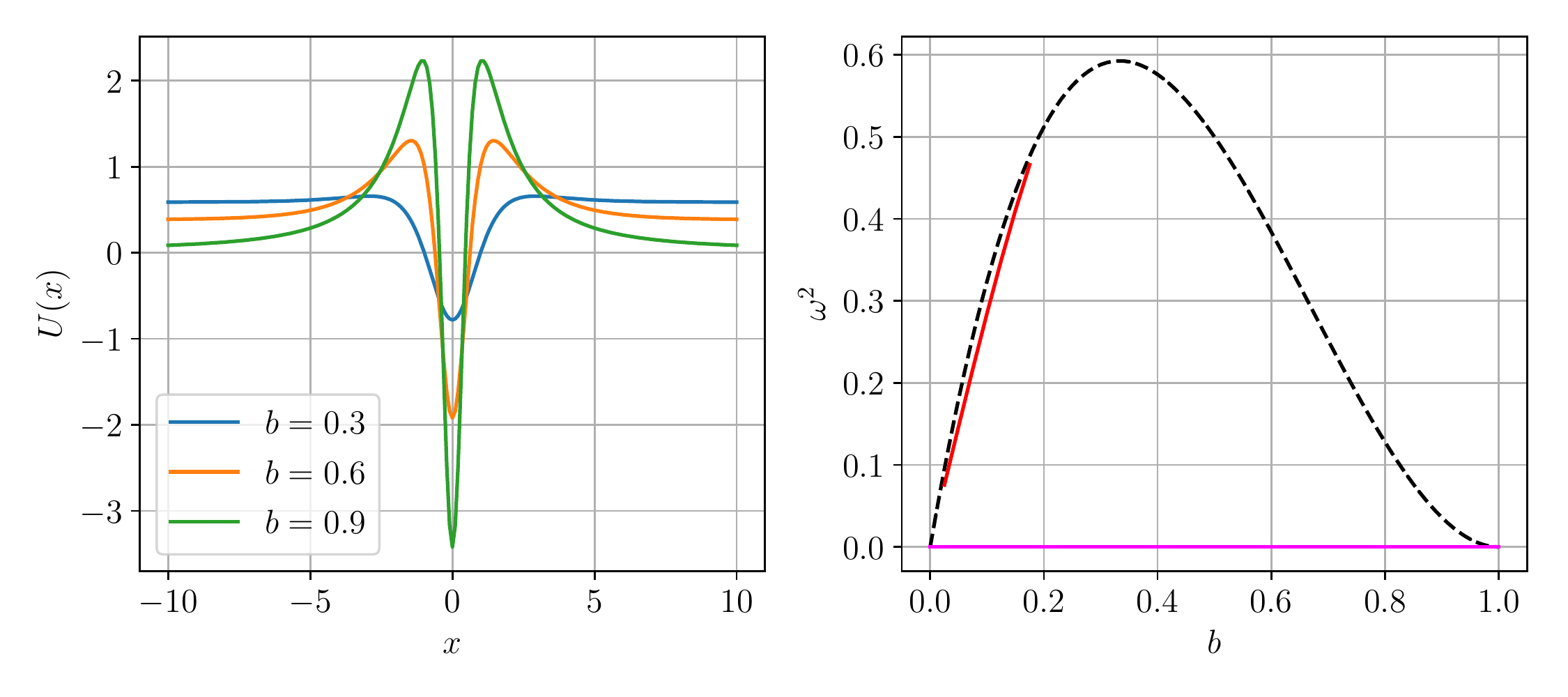}
    \end{subfigure}
    \begin{subfigure}{0.85\textwidth} 
    \includegraphics[width=\textwidth]{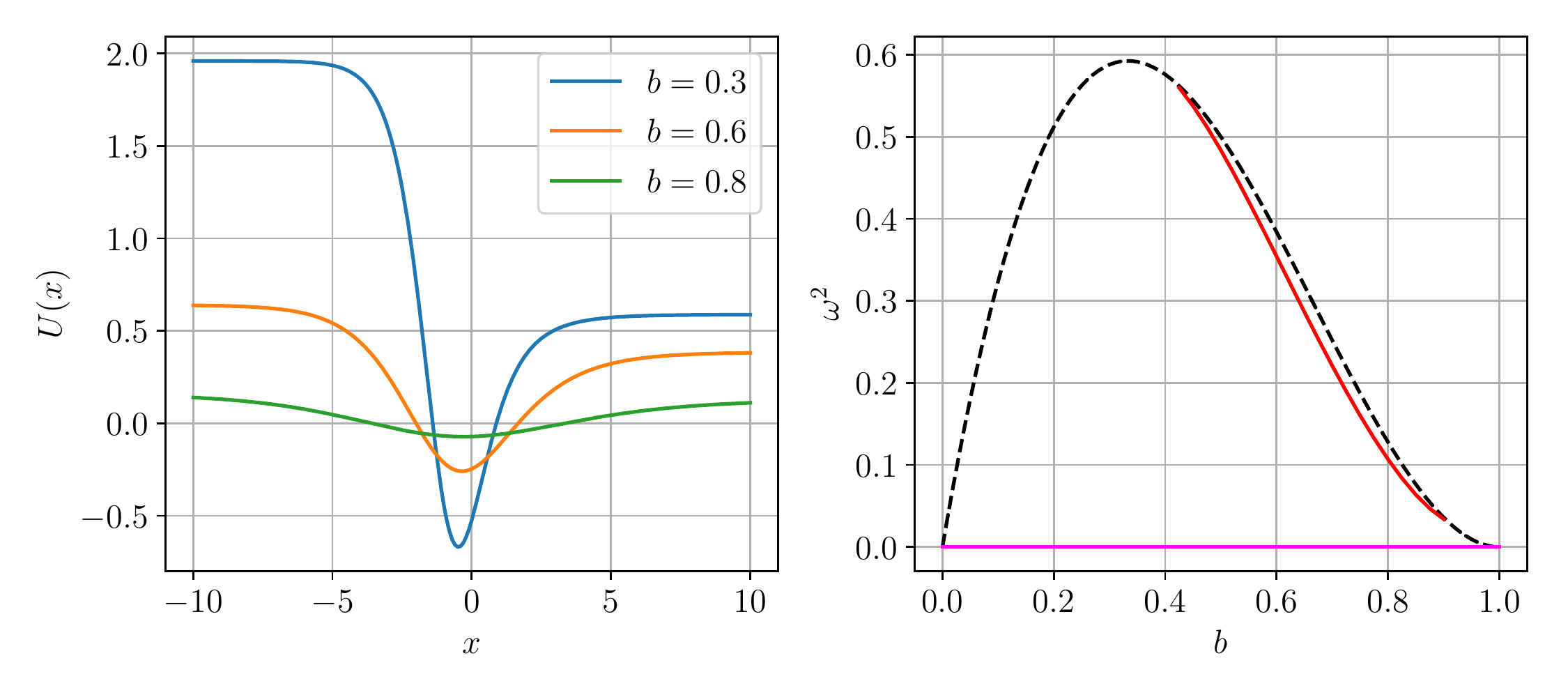}
    \end{subfigure}
    \caption{(Left) Effective potential for several values of $b$. (Right) The corresponding discrete spectra as a function of $b$. The top and bottom rows correspond to the inner and outer kinks, respectively. The zero mode frequency is shown in magenta and the first excited state is shown in red. The continuum spectrum starts at the black dashed line.}
    \label{fig:efpot1}
\end{figure}

The potential $U(x)$ and its spectrum for several values of $b$ are shown in Fig.~\ref{fig:efpot1}. The top row describes the effective potential for the symmetric inner kink $\phi_K^{(i)}$ with parameter $0<b<1$. For small $b$, the potential is a well, and it acquires a volcano shape as $b$ increases, consistent with the long-range character. 
The threshold energy which separates the discrete from the continuum modes is shown as a dashed line in the right panels. It is equal to the mass of the excitations around the vacua (mesons). For the inner kink, we find that $U_{\text{th}}(b)=4b(1-b)^2$. It is interesting to take a look at two limiting values. We have that $U_{\text{th}}(1)=0$ because the tail becomes long-range. Also $U_{\text{th}}(0)=0$, but for a different reason. This time, it occurs because the inner kink becomes increasingly small as $b\to 0$. 
Finally, we find that the spectrum contains only the zero mode, except for a small range where there is a single vibrational mode with a frequency close to the threshold value.

In the lower panels, we show the potential $U(x)$ for an asymmetric outer (anti)kink, $\phi_K^{(o)}$. The two asymptotic values of the potential are
\begin{equation}
    U(x\to-\infty)=4(1-b)^2,\quad U(x\to\infty)=4b(1-b)^2.
\end{equation}
The discrete spectrum is separated from the continuum at the smaller value, $U_{\text{th}}=4b(1-b)^2$. It is the same expression as in the previous case. Therefore, we obtain a vanishing threshold value again at the two limiting cases, that is, $U_{\text{th}}(0)=U_{\text{th}}(1)=0$. However, the picture is inverted for the outer kink, meaning that $b=0$ corresponds to the outer kink acquiring a long-range tail and $b=1$ to it becoming increasingly small. Again, the spectrum contains no vibrational modes or a single vibrational mode with a frequency close to the threshold of the continuous spectrum.

\section{Results}
\label{sec:Results}

We performed numerical evaluations of kink-antikink collisions. It is achieved by integrating the equations of motions starting with the following initial conditions
\begin{equation}
    \phi(x,t)=\phi_K(\gamma(x+X_0-v_it))+\phi_{\bar{K}}(\gamma(x-X_0+v_it))-C,
\end{equation}
where $X_0=12.5$ is the kink's initial position and $v_i$ is the absolute value of the initial velocity. This is known as the additive ansatz, and the constant $C$ is needed to adjust its limiting values. This expression approximates the system's dynamics well when the kinks are far away. Of course, evolution does not obey this superposition anymore as the kinks start to overlap. 

We will discuss all three existing types of kink-antikink collision. The first one is between inner kinks. Without loss of generality, we can consider only kink-antikink collisions. In the outer sector, one can have either kink-antikink or antikink-kink collisions because the kinks are asymmetric. To avoid ambiguities when referring to them, we will always start our simulations with outer kinks which interpolate between vacua with positive values of $\phi$.

The integration methods are standard. We discretize space in the interval $-400.0<x<400.0$ separated by 8192 equally spaced grid points and with periodic boundary conditions. The spatial derivatives are approximated by a Fourier spectral method, and the resulting set of ordinary differential equations are integrated using a fifth-order Runge-Kutta method with adaptive time step and error control. To avoid returning radiation from the boundaries, we include damping in the regions $x<-380.0$ and $x>380.0$. The damping is multiplied by a bump function \footnote{The bump function was obtained by shifting the center of the following expression to the boundaries $$B(x)=\begin{cases}5\exp\left(1-\frac{20^2}{20^2-x^2}\right),&-20<x<20,\\0, &\text{otherwise}.\end{cases}.$$} with a maximum value at the boundaries $x=\pm 400.0$. We measure the error in our numerical method by calculating the total energy conservation when the damping is turned off. The maximum relative error in our simulations is of order $10^{-5}$.

Now, care must be taken when simulating kinks with long-range tails because the additive ansatz is not a good approximation anymore. The same is true when the kinks become very small. As described above, the two phenomena appear in the limits $b\to 0$ and $b\to 1$. Here, we employ the method developed in \cite{campos2021interaction}, adding an extra minimization level on the one developed in \cite{christov2019long} to initialize such systems. In short, the method consists of minimizing the configuration obtained by the additive ansatz in order for the field $\phi$ to obey the static kink equation as closely as possible and the velocity field $\dot{\phi}$ to obey the zero mode equation as close as possible. This is performed via a nonlinear least-square minimization. In order to the minimization procedure be computationally viable we reduced the box to $-100.0<x<100.0$ separated by 2048 grid points, while the damping was included in the regions $x<-80.0$ and $x>80.0$.

In Ref.~\cite{khare2022kink}, the authors raised doubt on the efficiency of the minimization method in \cite{campos2021interaction}. Here, we employed the method extensively, showing that it is viable for large-scale computer simulations. Therefore, we give further evidence that the method is highly computationally efficient.

Below we will use colors to describe the field values. To make the comparison between graphs easier, we will use the same color scheme in all graphs in the same subsection. However, the color schemes in different subsections may be different.

\subsection{Inner sector kink-antikink collision}

The field evolution in this sector corresponds to a collision between an inner kink and an inner antikink. An important phenomenon that we will observe is kinks changing sector. It means that when the kinks get in contact, they do not bounce. Instead, they continue moving and form a pair of kinks in the neighboring sector, the outer kinks. This is a well-known phenomenon that happens, for instance, in the sine-Gordon kink-antikink collisions.

\begin{figure}
    \centering
    \includegraphics[width=0.85\textwidth]{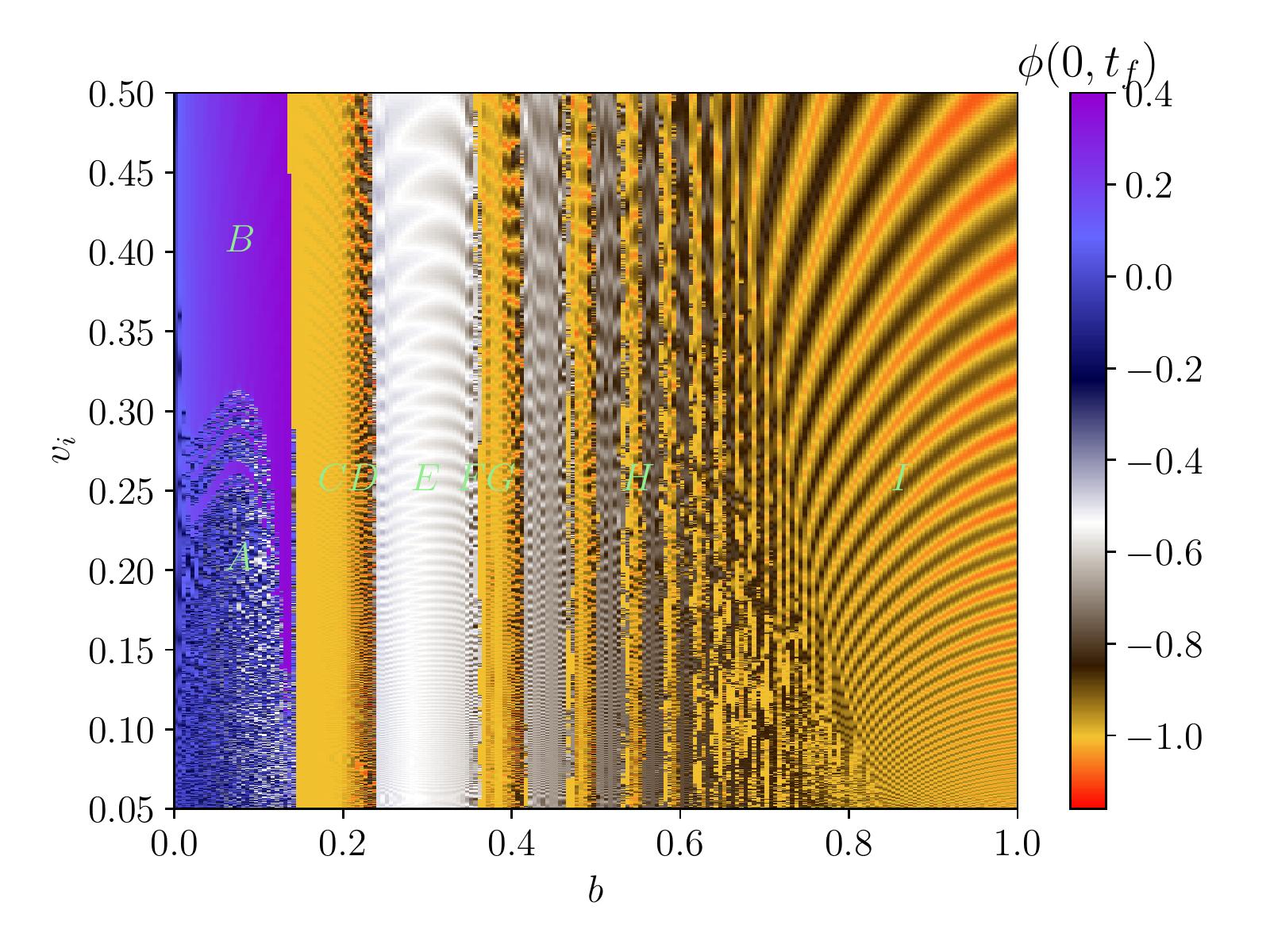}
    \caption{Final value of $\phi$ at the collision center as a function of $b$ and $v_i$. We are considering kink-antikink collisions between inner kinks.}
    \label{fig:mat-inner}
\end{figure}

\begin{figure}
    \centering
    \begin{subfigure}{\textwidth} 
    \includegraphics[width=\textwidth]{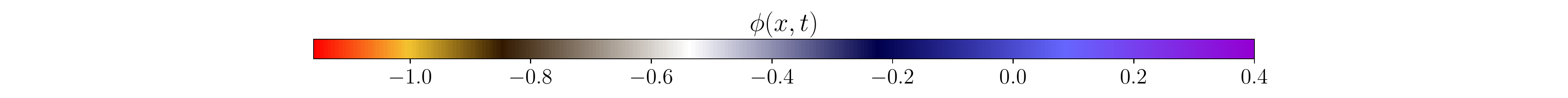}
    \end{subfigure}
    \begin{subfigure}{0.32\textwidth}
    \caption{$b=0.05$, $v_i=0.2$.}
    \includegraphics[width=\textwidth]{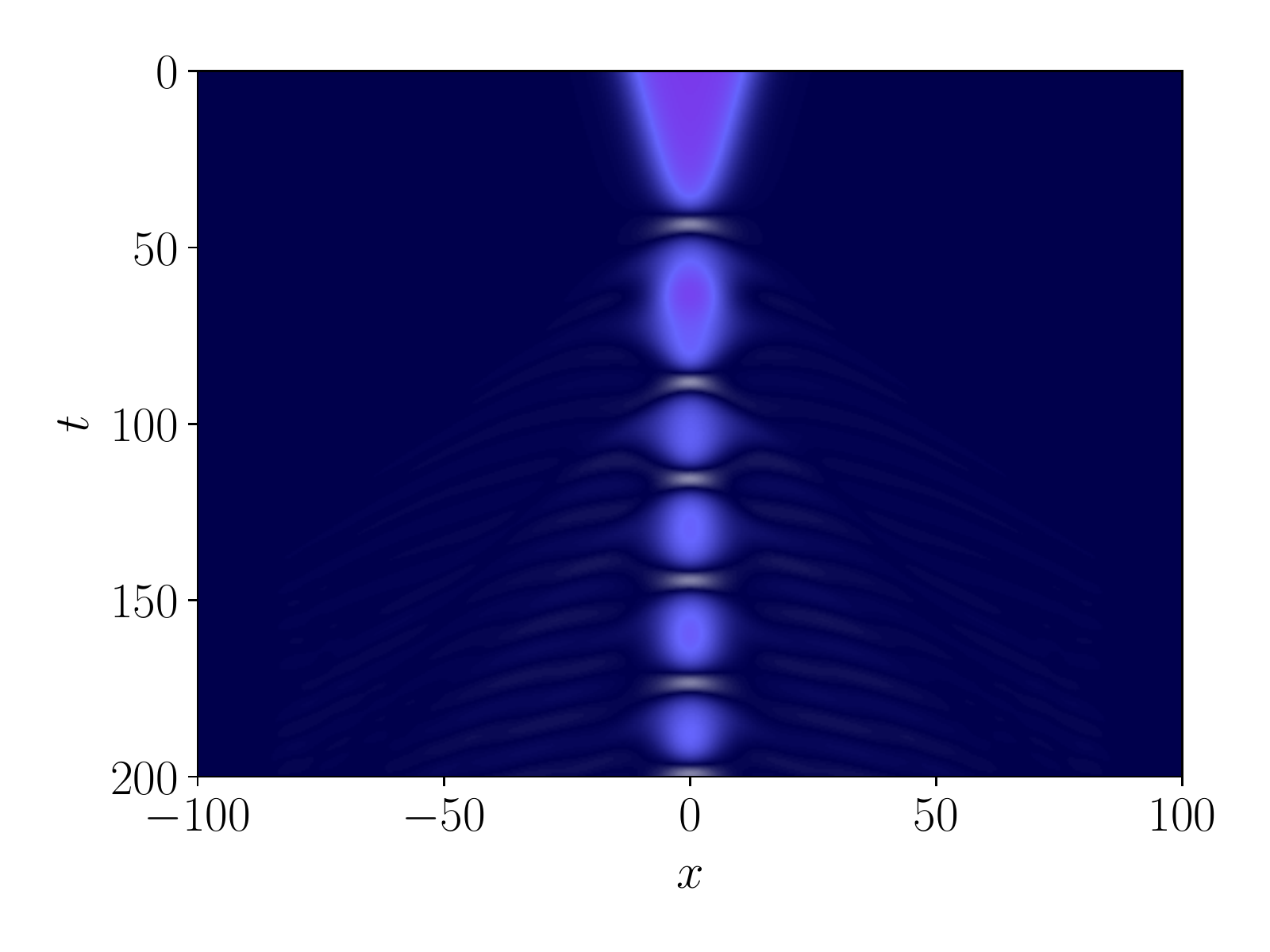}
    \end{subfigure}
    \begin{subfigure}{0.32\textwidth} 
    \caption{$b=0.1$, $v_i=0.4$.}
    \includegraphics[width=\textwidth]{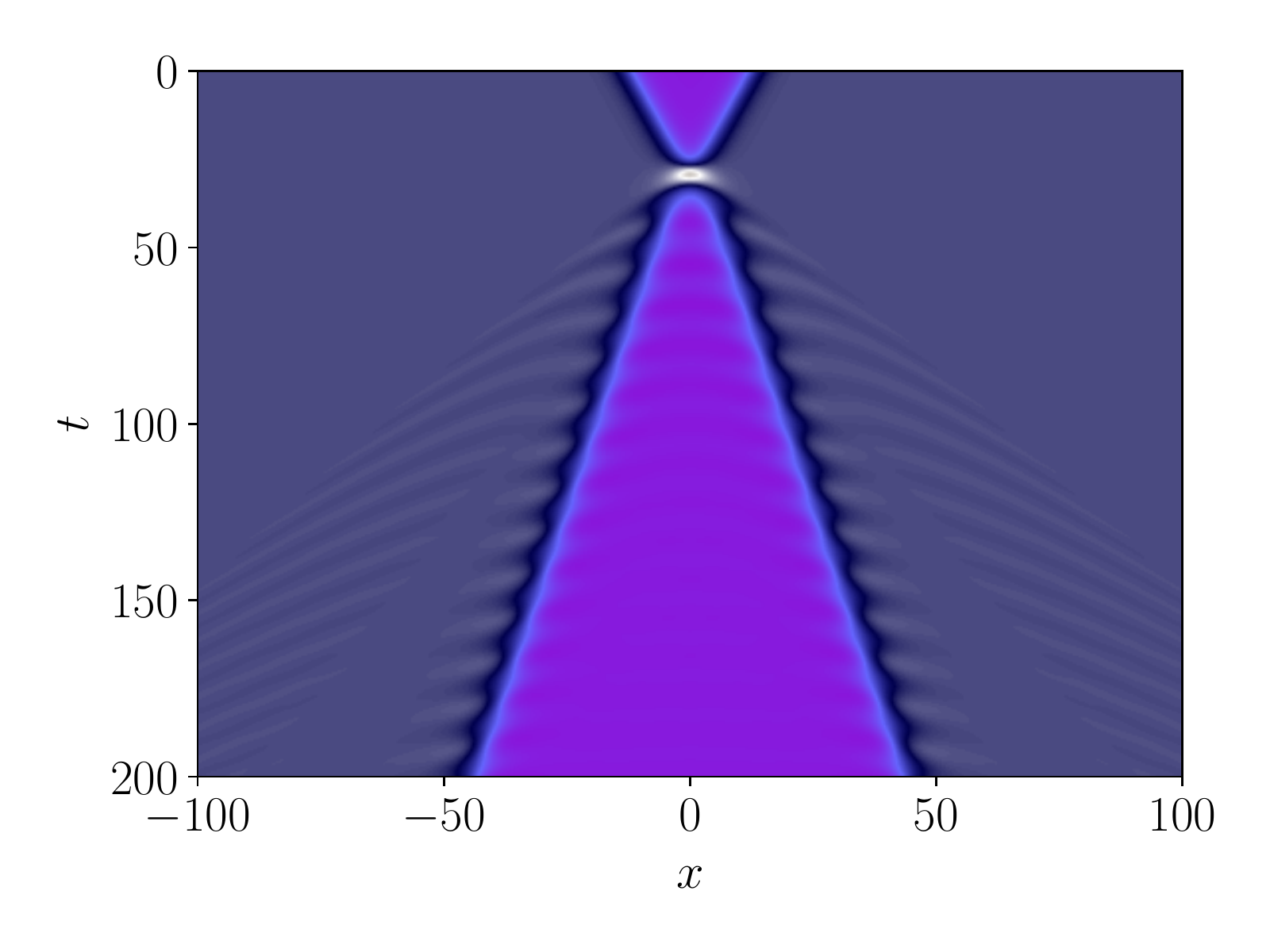}
    \end{subfigure}
    \begin{subfigure}{0.32\textwidth} 
    \caption{$b=0.17$, $v_i=0.2$.}
    \includegraphics[width=\textwidth]{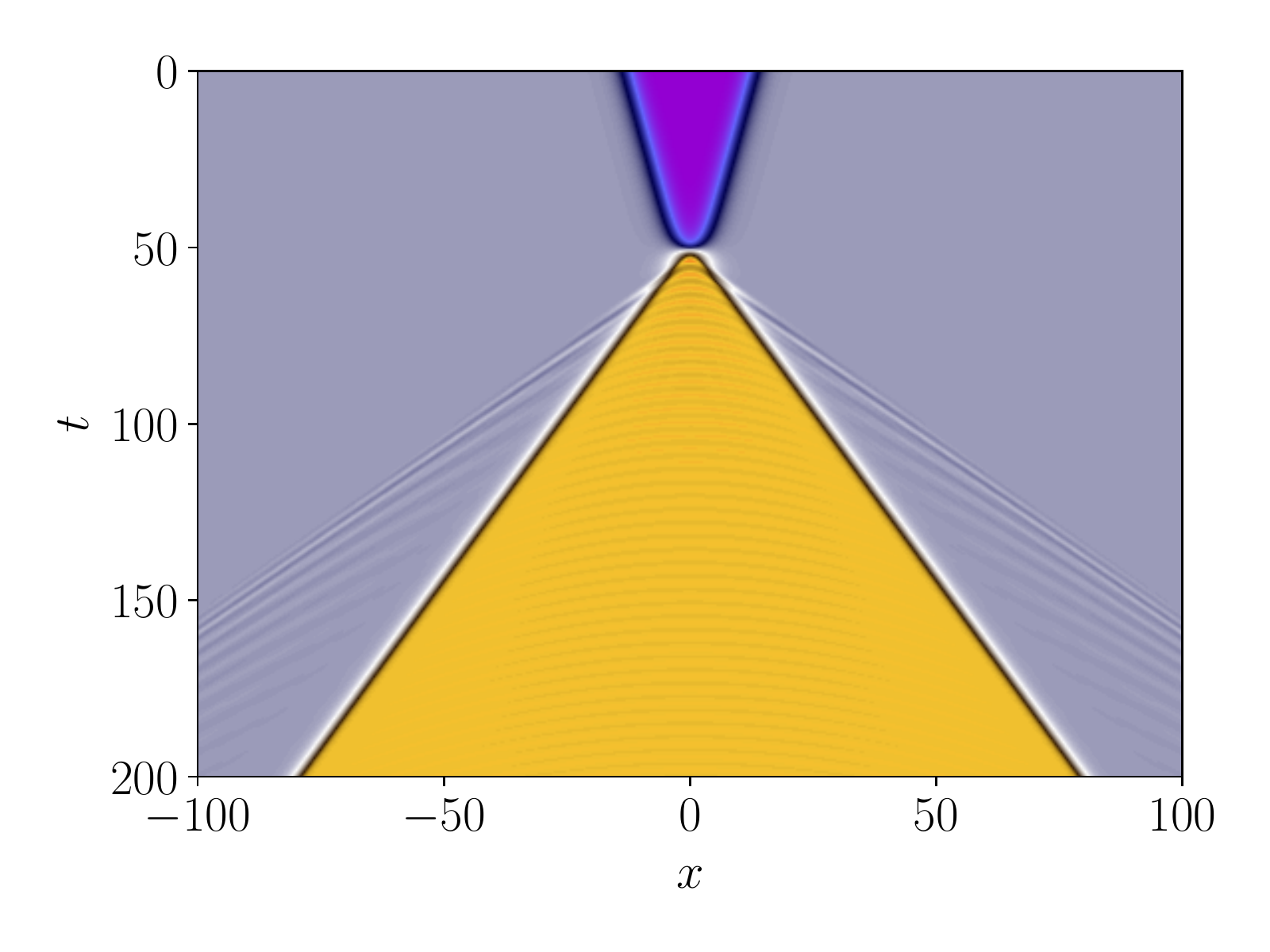}
    \end{subfigure}
    \begin{subfigure}{0.32\textwidth} 
    \caption{$b=0.23$, $v_i=0.2$.}
    \includegraphics[width=\textwidth]{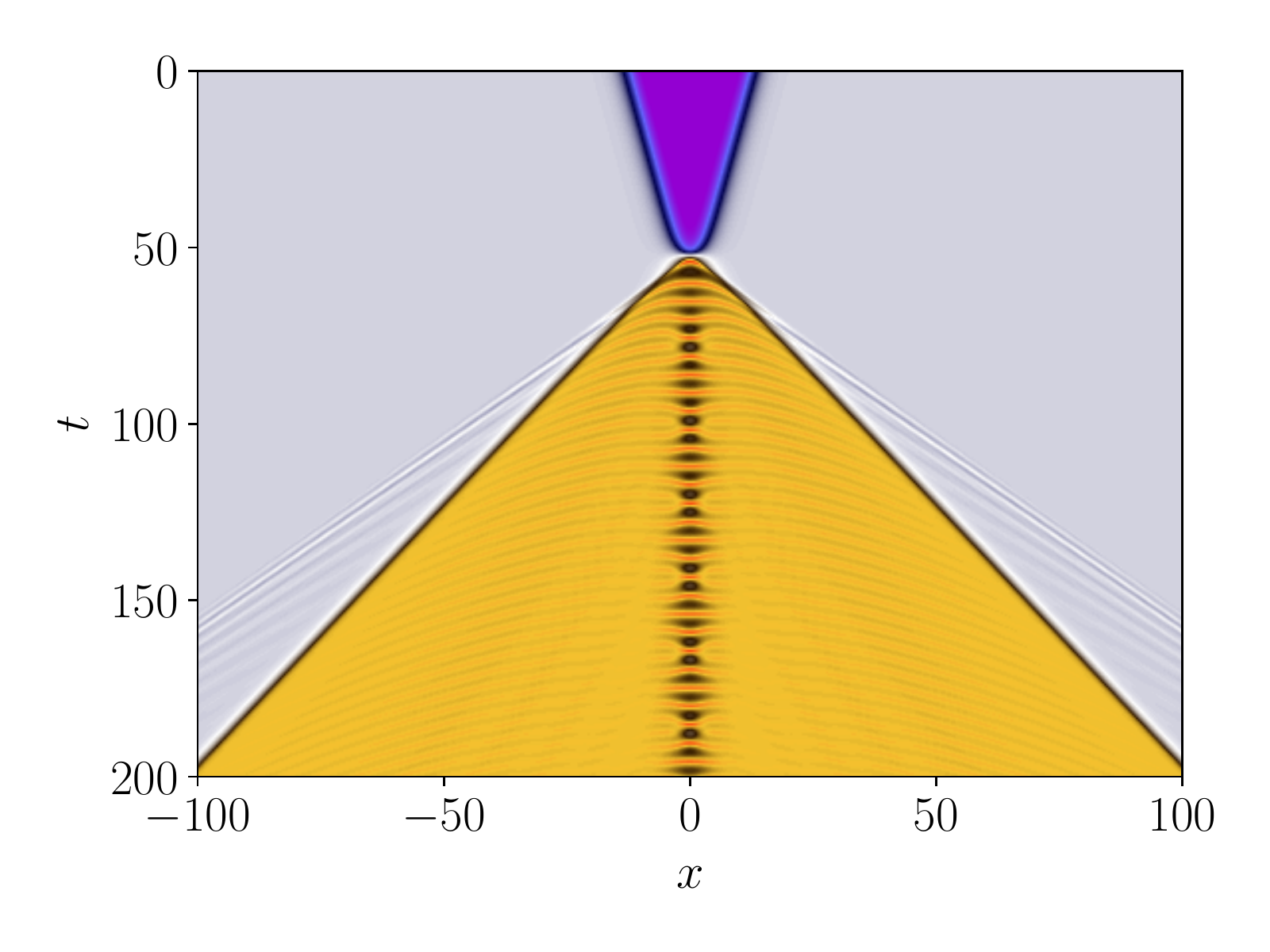}
    \end{subfigure}
    \begin{subfigure}{0.32\textwidth} 
    \caption{$b=0.26$, $v_i=0.2$.}
    \includegraphics[width=\textwidth]{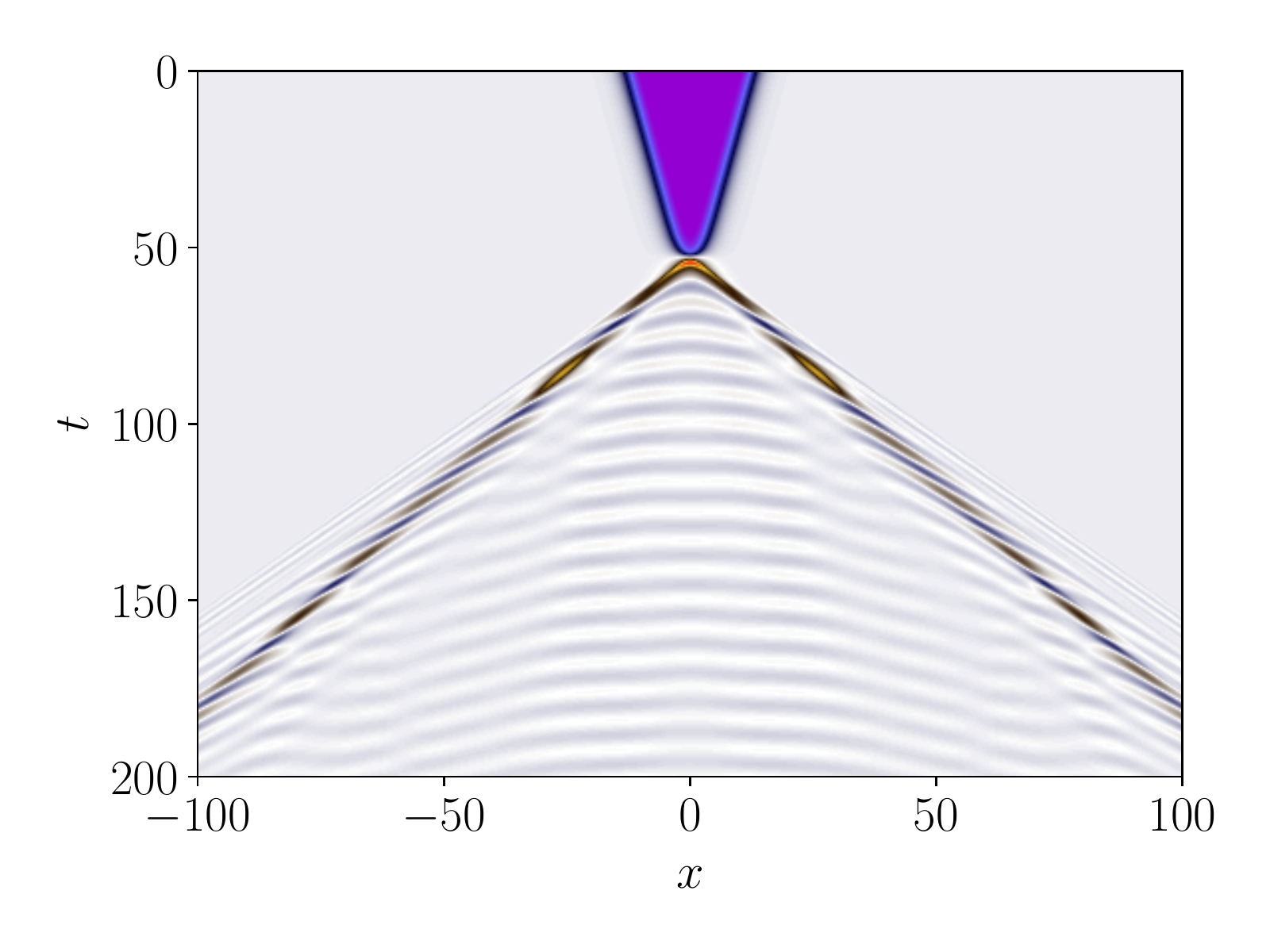}
    \end{subfigure}
    \begin{subfigure}{0.32\textwidth} 
    \caption{$b=0.358$, $v_i=0.2$.}
    \includegraphics[width=\textwidth]{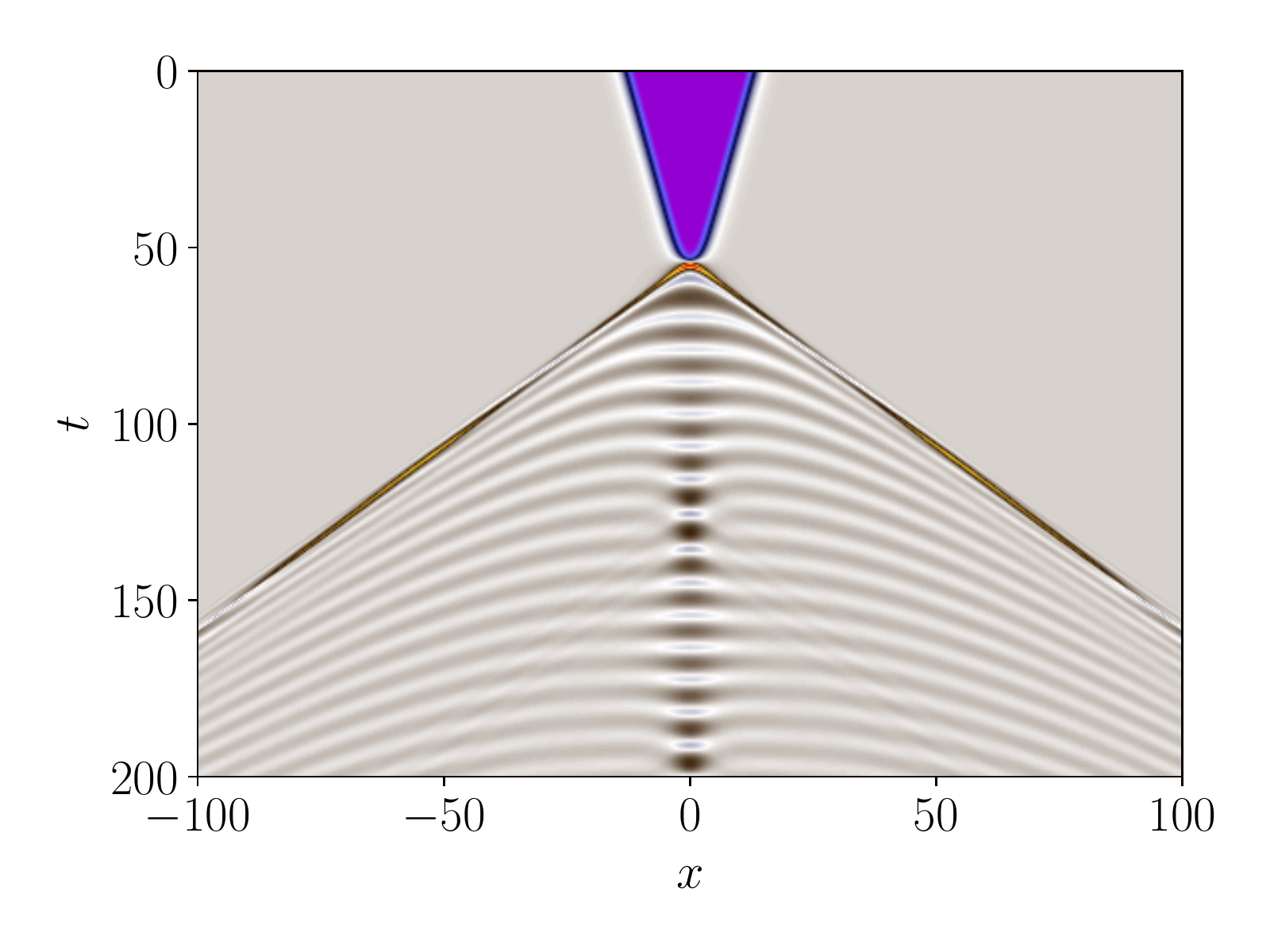}
    \end{subfigure}
    \begin{subfigure}{0.32\textwidth} 
    \caption{$b=0.37$, $v_i=0.2$.}
    \includegraphics[width=\textwidth]{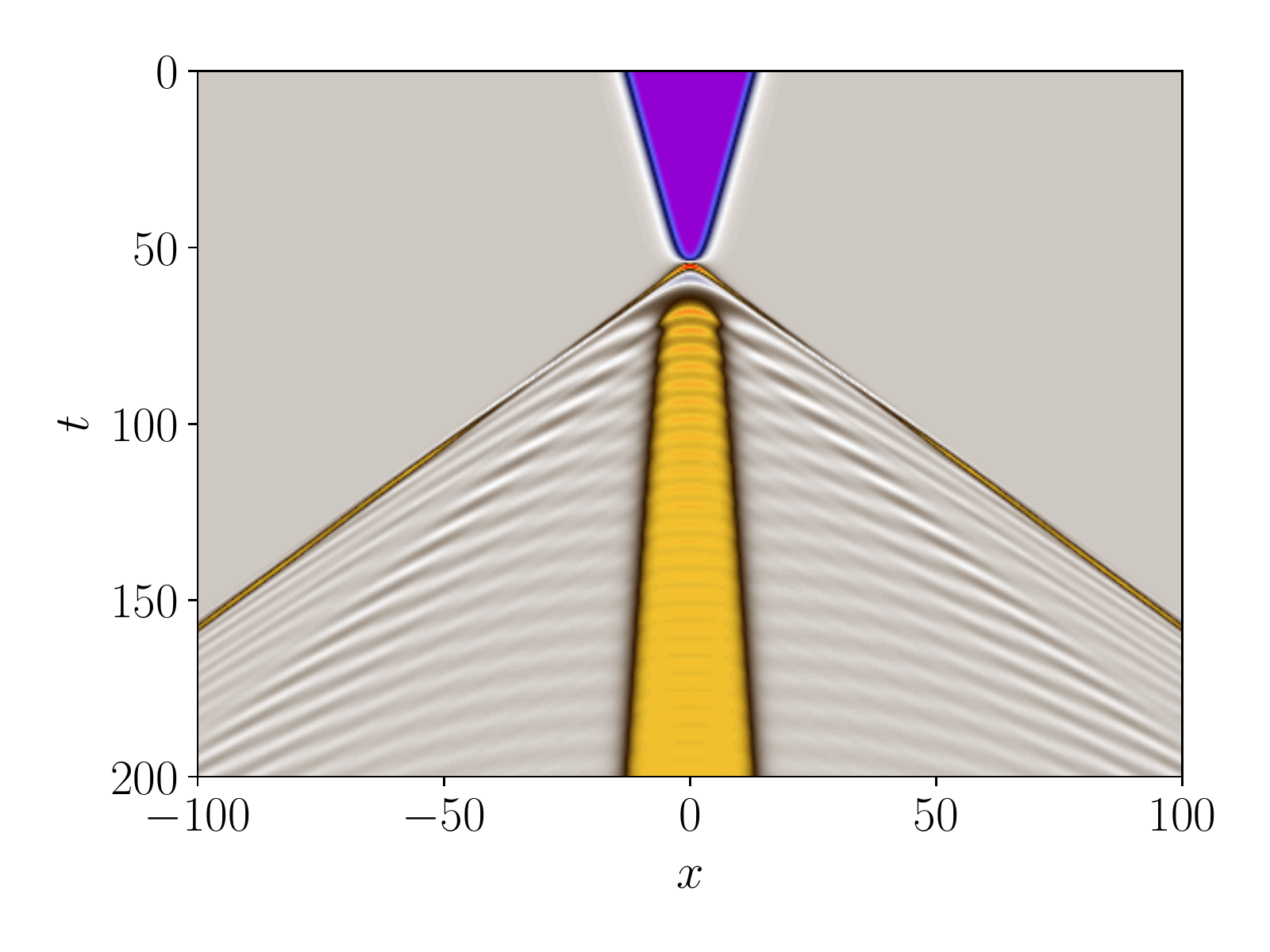}
    \end{subfigure}
    \begin{subfigure}{0.32\textwidth} 
    \caption{$b=0.55$, $v_i=0.2$.}
    \includegraphics[width=\textwidth]{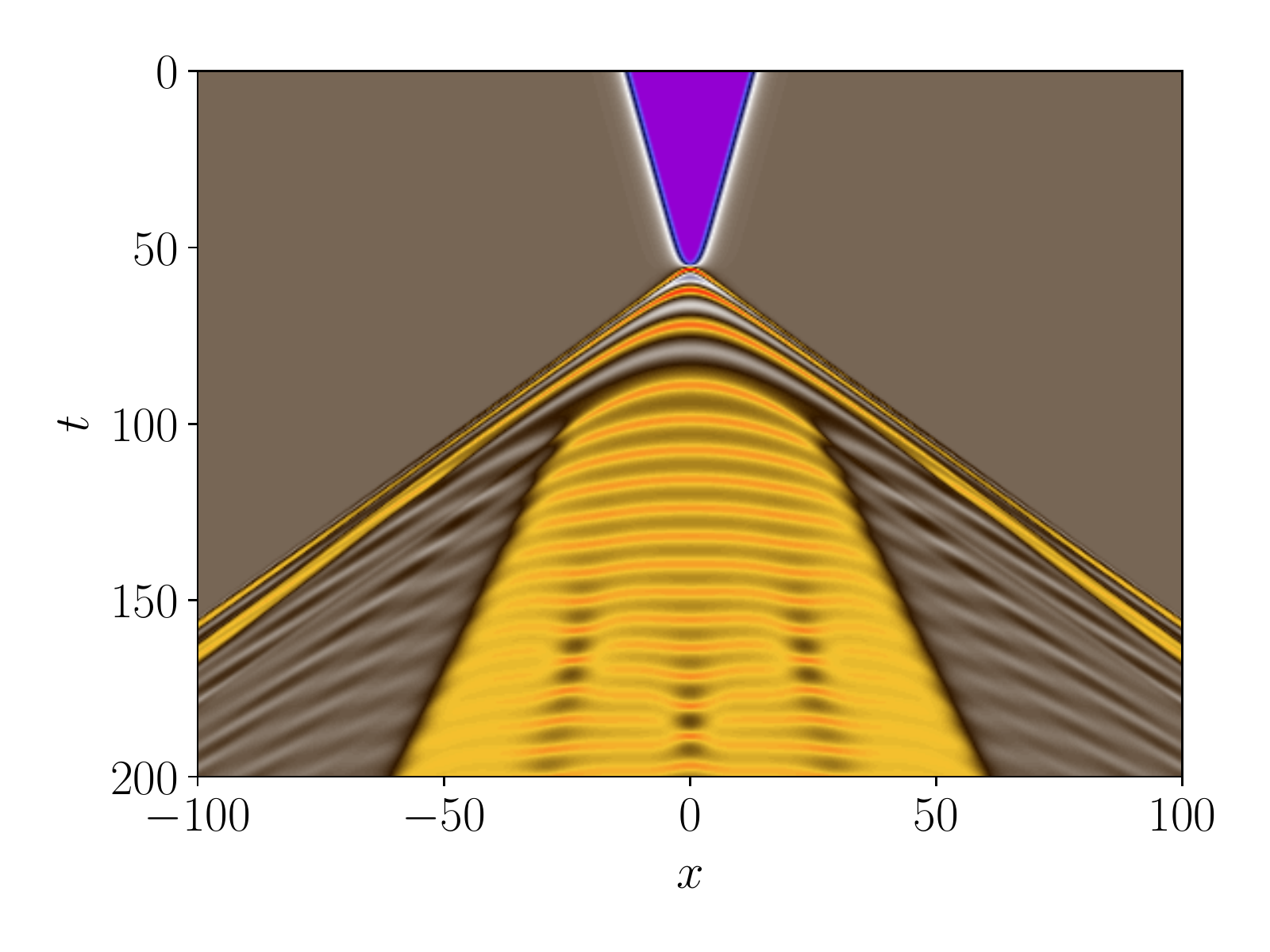}
    \end{subfigure}
    \begin{subfigure}{0.32\textwidth} 
    \caption{$b=0.82$, $v_i=0.2$.}
    \includegraphics[width=\textwidth]{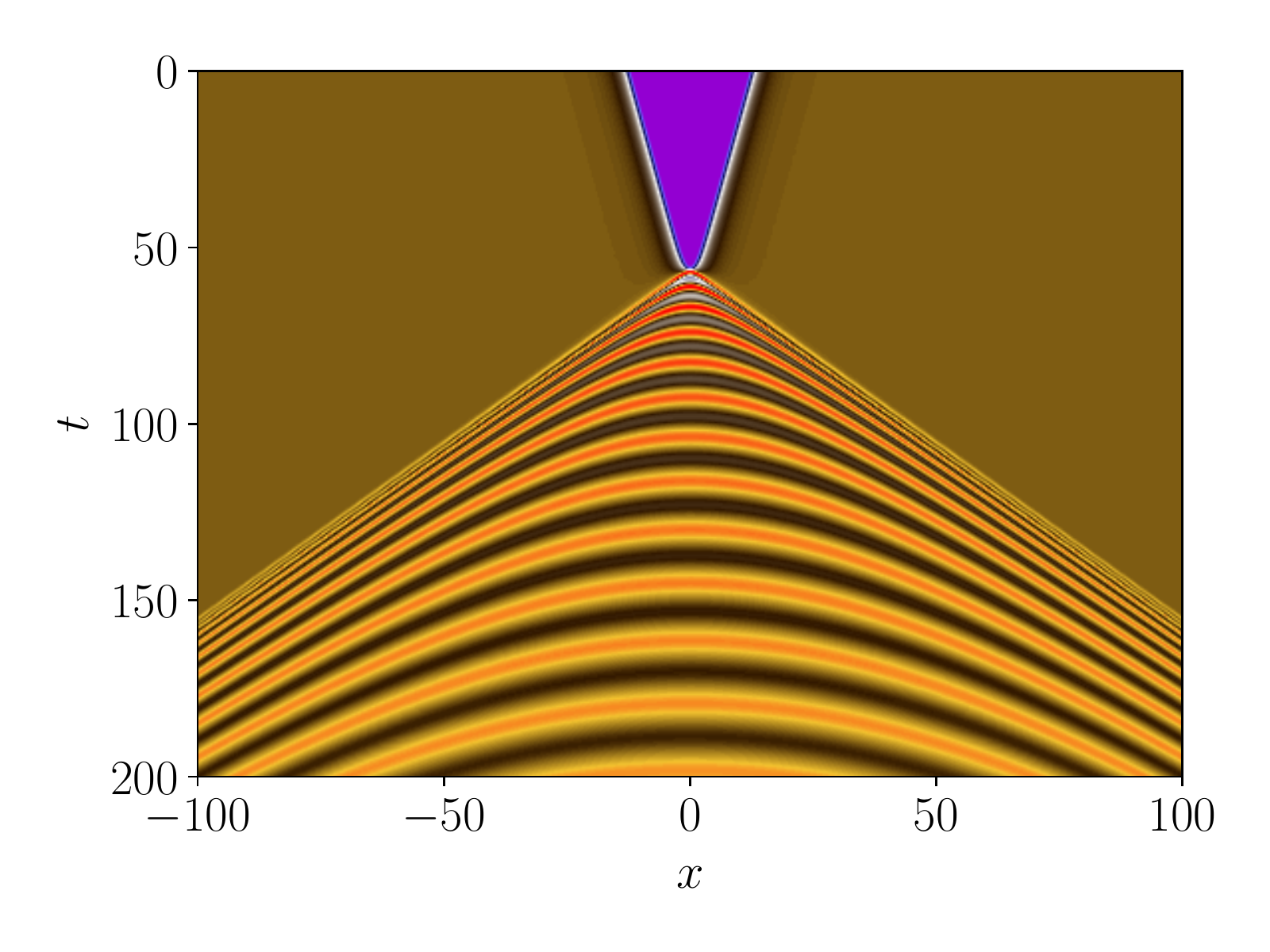}
    \end{subfigure}
    \caption{Evolution of $\phi(x,t)$ in spacetime for kink-antikink collisions between inner kinks.}
    \label{fig:field-inner}
\end{figure}

The system's behavior as a function of the parameter $b$ and the initial velocity $v_i$ can be summarized in Fig.~\ref{fig:mat-inner}. The color corresponds to the field's value at the collision center at a time $t_f=60.0/v_i$. It is possible to distinguish several different regions. Some of them are listed below:
\begin{itemize}
    \item Region $A$: The kinks form a slowly decaying bound state after colliding. Such a state is called a bion, which is illustrated in Fig.~\ref{fig:field-inner}(a).
    \item Region $B$: The kinks collide and then reflect as illustrated in Fig.~\ref{fig:field-inner}(b).
    \item Region $C$: The kinks collide, change sectors, and then separate. This case is illustrated in Fig.~\ref{fig:field-inner}(c).
    \item Region $D$: Same as $C$, but now an oscillating pulse is formed in the center. Fig.~\ref{fig:field-inner}(d) illustrates this case.
    \item Region $E$: The kinks collide and change sectors. Moreover, an extra pair of outer kink is formed. This leads to four outer kinks in the final state. Fig.~\ref{fig:field-inner}(e) illustrates this case. In the case presented in the figure, the four outer kinks form two bions. A significant amount of radiation is also present.
    \item Region $F$: The kinks change sector and create an extra pair of outer kinks. Moreover, an oscillating pulse is formed at the center. Fig.~\ref{fig:field-inner}(f) illustrates this case.
    \item Region $G$: The kinks collide, forming three pairs of outer kinks in the final state. An example is given in Fig.~\ref{fig:field-inner}(g).
    \item Region $H$: The kinks collide and form seven pairs of outer kinks in the final state. An example is given in Fig.~\ref{fig:field-inner}(h).
    \item Region $I$: In the limit the outer kinks are very light, many kink-antikink pairs are formed. As the kinks are also very small, they form a pattern similar to the one in radiation emission. Fig.~\ref{fig:field-inner}(i) illustrates this case.
\end{itemize}

A similar behavior observed in regions $C$ to $H$ was previously reported in Ref.~\cite{simas2020solitary} when analyzing the double sine-Gordon model. We see that the outer kink mass decreases with $b$, creating more pairs as the mass vanishes for $b=1$. The evidence is $\phi(0,t_f)$ changing color from white to yellow. So every time a new kink-antikink pair is created, the vacuum at the center needs to be adjusted accordingly, indicated by the color change. Moreover, Our numerical simulations show that when the creation of a pair is almost energetically favored, oscillating pulses appear. Increasing further the initial energy, the oscillating pulse turns into a kink-antikink pair. This result is in agreement with Ref.~\cite{simas2020solitary}. Moreover, the extra kink-antikink pairs may form bound states among them. However, in the present case, the bound structures are less stable than in Ref.~\cite{simas2020solitary}, meaning that they are more likely to decay into radiation.

As $b$ approaches one, all the extra kink pairs create a radiation pattern. Furthermore, we see in Fig.~\ref{fig:mat-inner} that this pattern repeats at least up to the largest exhibited value, $v_i=0.5$. Hence, to overcome this effect, a large initial velocity is needed. For $b=1$, the outer kinks cease to exist, and the inner kink acquires long-range tails at both sides. This case was studied in Ref.~\cite{campos2021interaction}, and the same pattern, as well as an ultrarelativistic critical velocity, was observed. Therefore, we found how the annihilation into radiation is achieved as two quadratic minima merge to form a quartic one. It is through the creation of many small kink-antikink pairs.

In Ref.~\cite{khare2022kink}, the authors also raised doubt on the claims presented in Ref.~\cite{campos2021interaction}. The claims were that the kinks decay into radiation without bion formation and possess an ultrarelativistic critical velocity for the kinks with long-range tails on both sides. The key ingredient for the new behavior is a long-range tail in the direction that is not facing the opposing kink. Above, we presented a more detailed physical description of the phenomena, corroborating such claims. Interestingly, it has recently been reported a different scenario where complete kink-antikink annihilation can occur, which is the scattering of long-range kinks by a half-BPS preserving impurity \cite{adam2023moduli}.

In regions $A$ and $B$, the outer kinks are more massive than the inner ones. Hence, it is not possible to change sectors, and the kinks bounce back when they collide, resembling the $\phi^4$ model. There are also resonance windows at the boundary between the two regions, which is consistent with the presence of a vibrational mode. Moreover, the inner kinks become increasingly small as $b$ vanishes. In such a case, we observe that the critical velocity separating regions $A$ and $B$ sharply increases near this point. This pattern will also occur in the other cases considered below.

The simulations for $b\leq0.6$ were performed after minimizing the initial condition as described previously. Such a procedure was crucial to obtain correct results. In the limit $b\to1$, the minimization procedure gives correct results for $v>0.25$. However, there is no visible difference compared to non-minimized simulations. For $v<0.25$, our minimized algorithm runs into trouble because the box is too small and $t_f$ is too large. Note that the collision creates a large amount of radiation which travels at speed near $c=1$, see Fig.~\ref{fig:field-inner}(i). This creates a spurious effect from returning radiation that cannot be eliminated by the damping term at the boundaries. Therefore, we did not use minimized initial conditions in this region, but we ran several tests to ensure that the result is still physically correct.

\subsection{Outer sector antikink-kink collision}

In this section, we will discuss the field evolution of antikink-kink collisions in the outer sector. In such case, the field in the region between the kinks is at the inner vacuum $\phi=\sqrt{b}$. Among all cases, this is the simplest one. In this sector, the kink-antikink configuration allows delocalized modes due to the asymmetry of the isolated kinks. Such modes are excitations of the scalar field, which appear due to the potential well created by the antikink-kink pair. Moreover, there is also a localized vibrational mode. We will see shortly how these affect the resonance structure.

\begin{figure}
    \centering
    \includegraphics[width=0.85\textwidth]{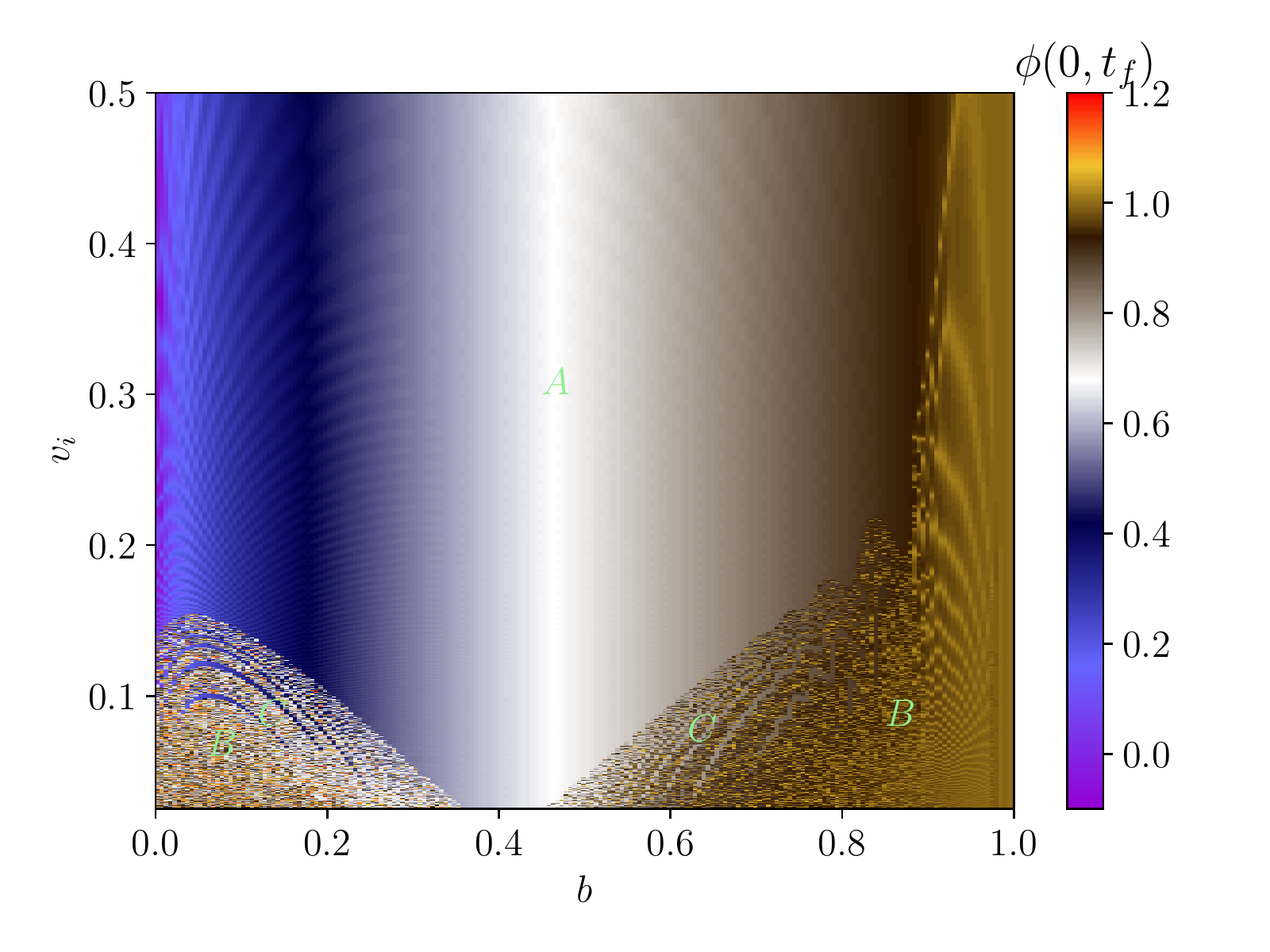}
    \caption{Final value of $\phi$ at the collision center as a function of $b$ and $v_i$. We are considering antikink-kink collisions between outer kinks.}
    \label{fig:mat-out-ak}
\end{figure} 

\begin{figure}
    \centering
    \begin{subfigure}{\textwidth} 
    \includegraphics[width=\textwidth]{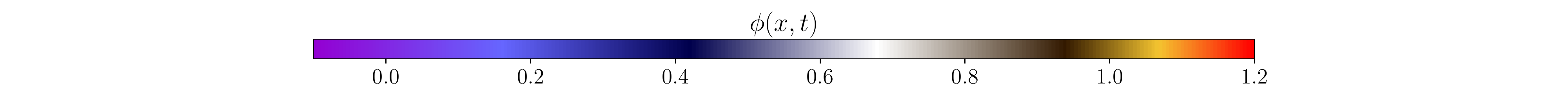}
    \end{subfigure}
    \begin{subfigure}{0.32\textwidth} 
    \caption{$b=0.45$, $v_i=0.3$.}
    \includegraphics[width=\textwidth]{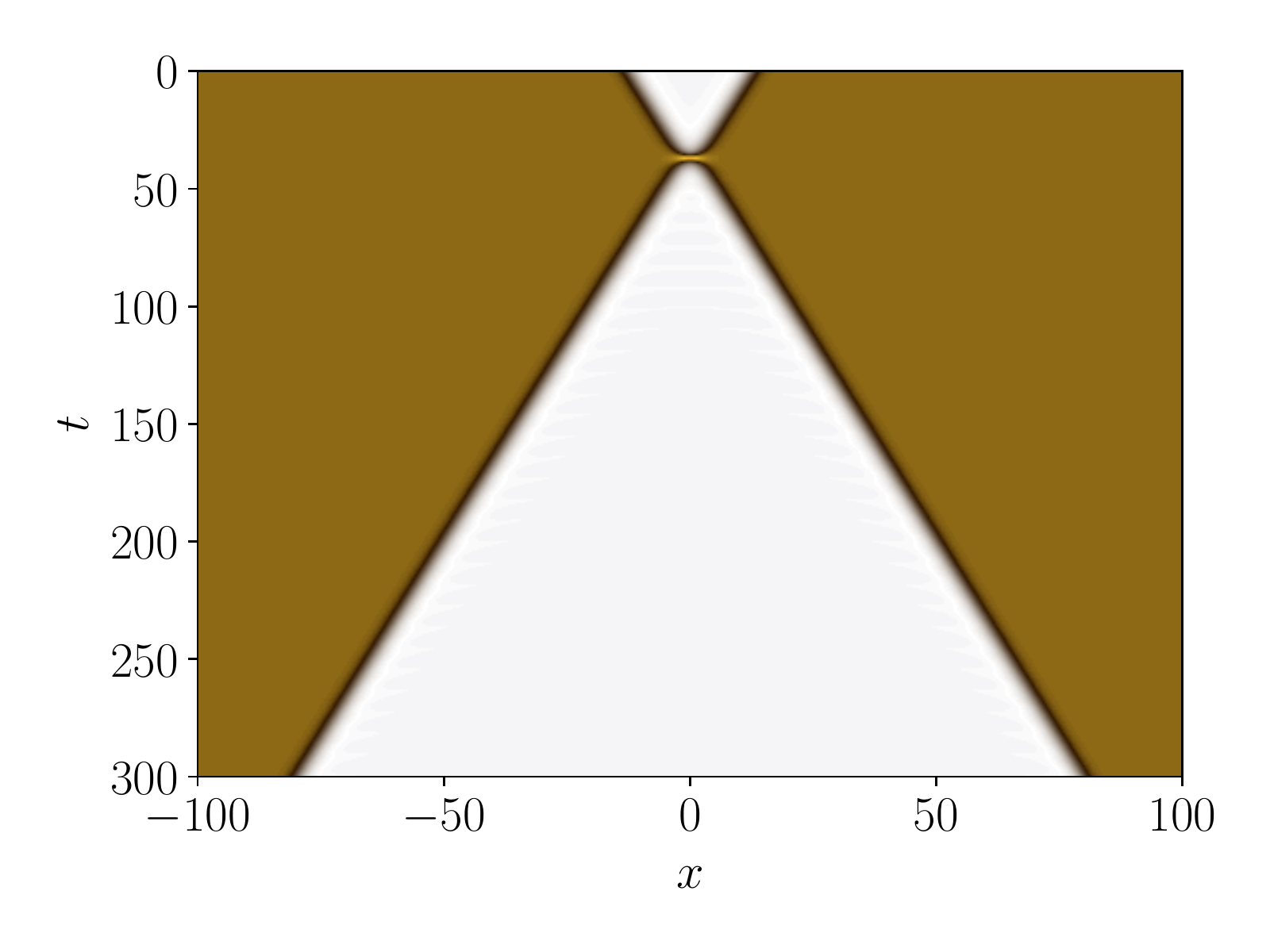}
    \end{subfigure}
    \begin{subfigure}{0.32\textwidth} 
    \caption{$b=0.07$, $v_i=0.06$.}
    \includegraphics[width=\textwidth]{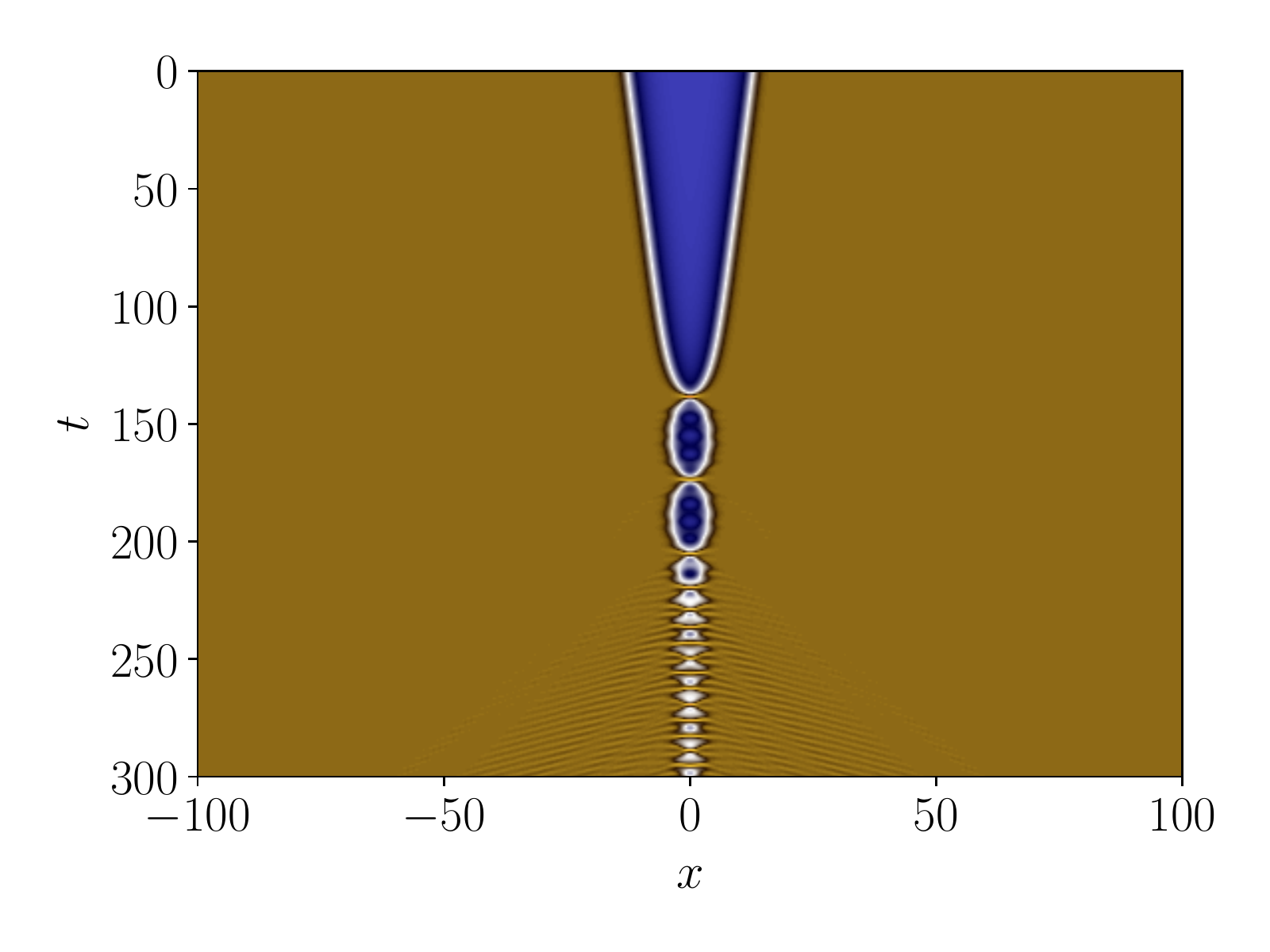}
    \end{subfigure}
    \begin{subfigure}{0.32\textwidth} 
    \caption{$b=0.15$, $v_i=0.09$.}
    \includegraphics[width=\textwidth]{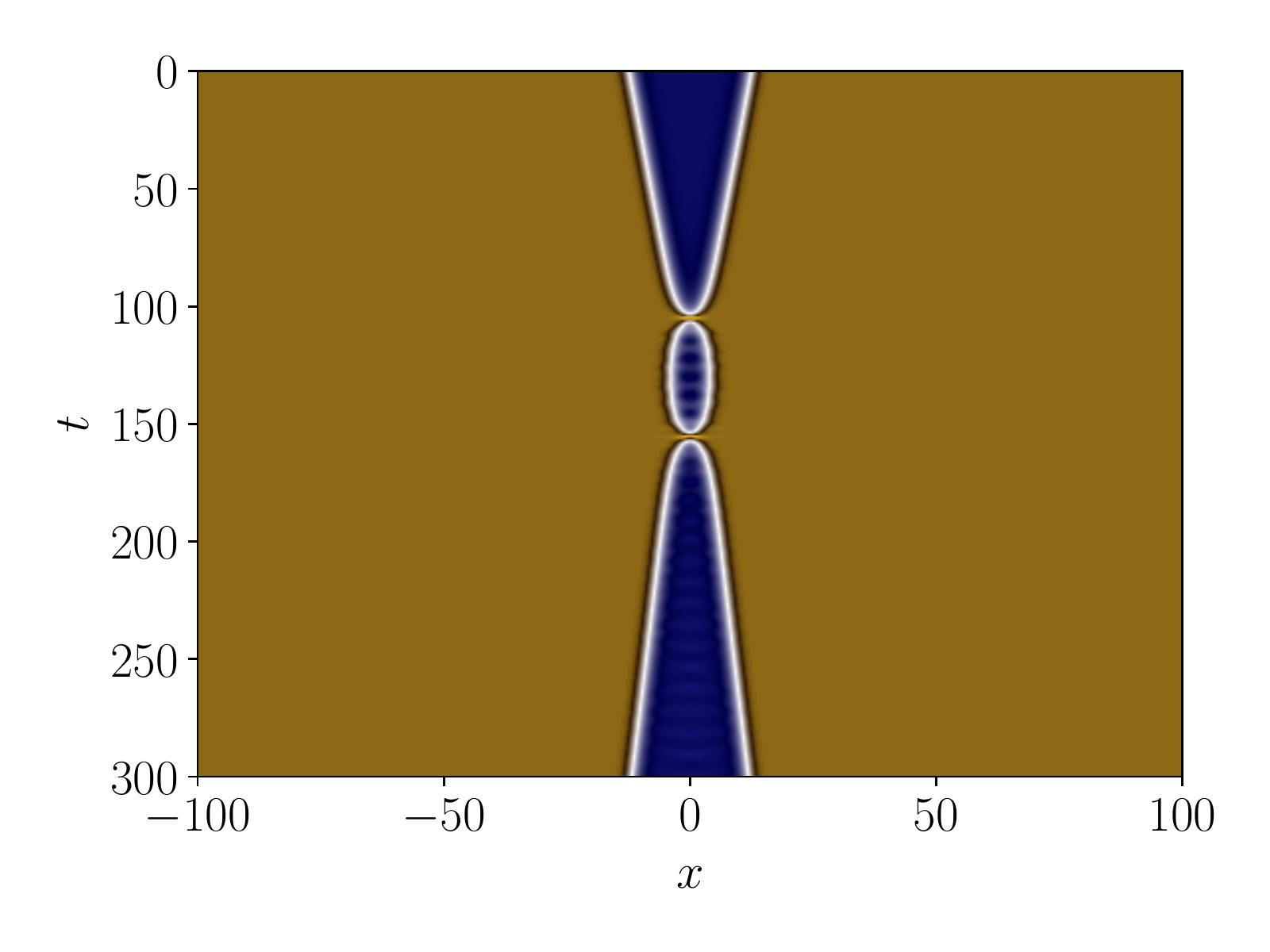}
    \end{subfigure}
    \begin{subfigure}{0.32\textwidth} 
    \caption{$b=0.12$, $v_i=0.089$.}
    \includegraphics[width=\textwidth]{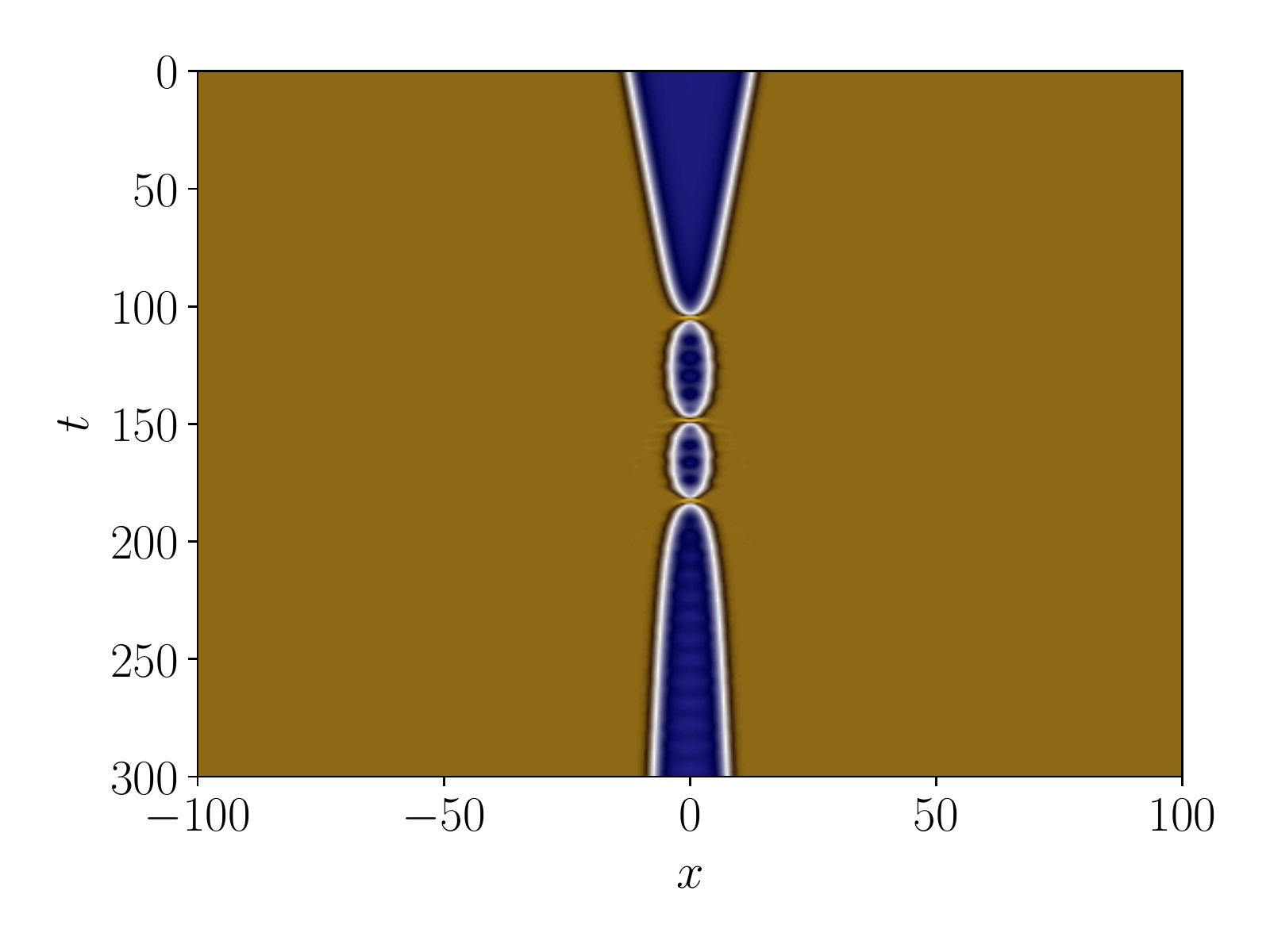}
    \end{subfigure}
    \begin{subfigure}{0.32\textwidth} 
    \caption{$b=0.66$, $v_i=0.074$.}
    \includegraphics[width=\textwidth]{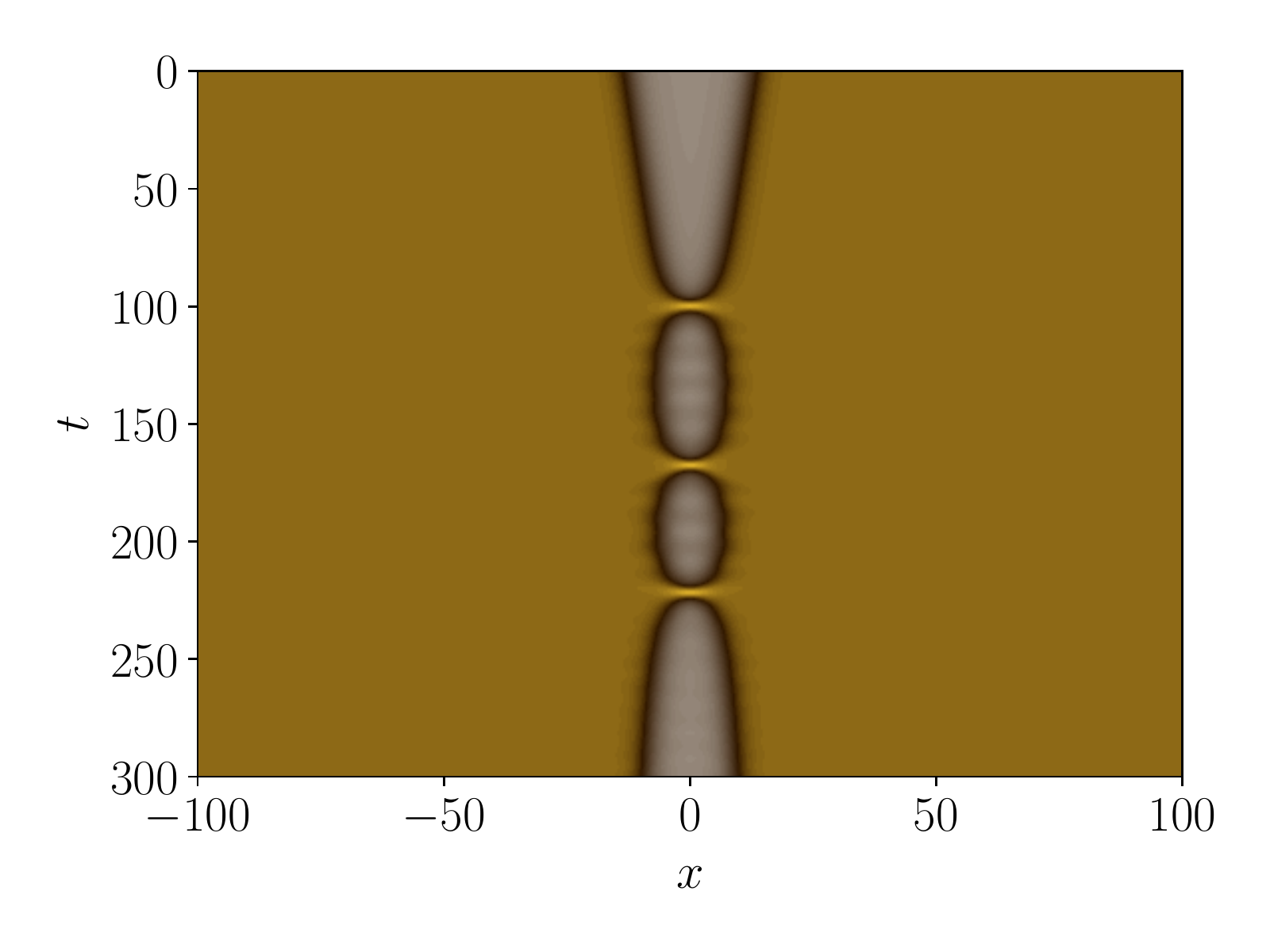}
    \end{subfigure}
    \begin{subfigure}{0.32\textwidth} 
    \caption{$b=0.125$, $v_i=0.109$.}
    \includegraphics[width=\textwidth]{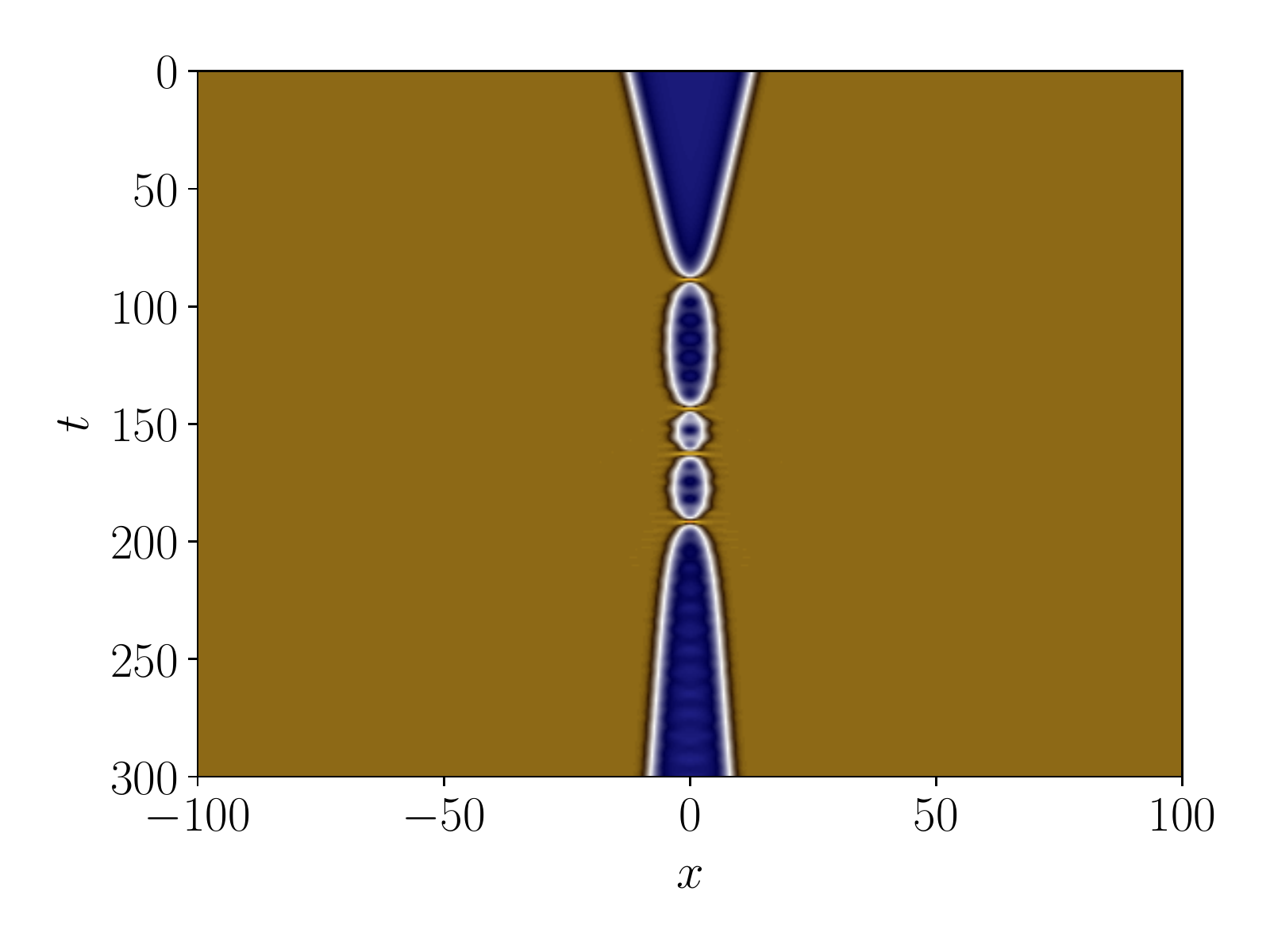}
    \end{subfigure}
    \caption{Evolution of $\phi(x,t)$ in spacetime for antikink-kink collisions between outer kinks.}
    \label{fig:field-out-ak}
\end{figure}

The system's behavior as a function of the parameter $b$ and the initial velocity $v_i$ is summarized in Fig.~\ref{fig:mat-out-ak}. It shows the field's value at the collision center at a time $t_f=60.0/v_i$. It is possible to distinguish between three different regions, which are listed below:
\begin{itemize}
    \item Region $A$: In the large smooth region above $B$ and $C$, the kinks collide and then reflect. The behavior is illustrated in Fig.~\ref{fig:field-out-ak}(a).
    \item Region $B$: The kinks form a bion, a state where the kink and the antikink are bound. The bion slowly decays to a trivial vacuum after a large number of collisions. Fig.~\ref{fig:field-out-ak}(b) illustrates this case.
    \item Region $C$: The kinks collide and then depending on the initial velocity form a bion or reflect after more than a single bounce. The latter are resonance windows. The two-, three- and four-bounce windows are illustrated in Fig.~\ref{fig:field-out-ak}(c),  Figs.~\ref{fig:field-out-ak}(d) and (e), and Fig.~\ref{fig:field-out-ak}(f), respectively. 
\end{itemize}

Interestingly, nothing special happens as we approach the limit $b=0$, where the tail facing the opposing kink is long-range. In other words, the system's resonance structure at $b=0$ is qualitatively the same as that of systems with exponential tails, which also possess delocalized modes. This is expected from the analysis in \cite{christov2021kink}.

The critical velocity near $b=0.4$ becomes less than $0.025$ and is not visible in Fig.~\ref{fig:mat-out-ak}. Due to the kink's asymmetry, delocalized vibrational modes create resonance windows for $b<0.3$. This is the region where the kinks are mostly asymmetrical. However, not all values of $b$ in the interval $[0,1]$ exhibit resonance windows. It is possible to see that the windows are suppressed near $b=0.3$. By increasing the parameter $b$, a vibrational mode appears, and the resonance structure is recovered. Increasing even further, the vibrational mode and the resonance windows cease to exist again.

In region $C$, the set of resonance windows form a quasi-fractal structure, meaning that $(n+1)$-bounce windows accumulate near the edge of $n$-bounce windows. To illustrate this phenomenon, examples of three and four-bounce resonance windows are given in Fig.~\ref{fig:field-out-ak}. Moreover, we also found higher-bounce windows in kink-antikink collisions in the inner and outer sectors, evidencing that they also form a quasi-fractal structure.


Another effect observed in Fig.~\ref{fig:mat-out-ak} is that the kinks become increasingly small and the critical velocity increasingly large as $b$ approaches one. Moreover, we should mention that the simulations for $b<0.12$ and $b\geq0.88$ were performed after minimizing the initial condition as described previously. Such a procedure was crucial to obtain correct results.

\subsection{Outer sector kink-antikink collision}

The field evolution in this sector corresponds to a kink-antikink collision between outer kinks. The kinks are initially separated by the vacuum at $\phi=1$. The main differences with the previous case are that the kink's asymmetry character does not allow delocalized modes' presence, and the outer kink-antikink pair can change sector after a collision and turn into an antikink-kink pair in the inner sector.

\begin{figure}
    \centering
    \includegraphics[width=0.85\textwidth]{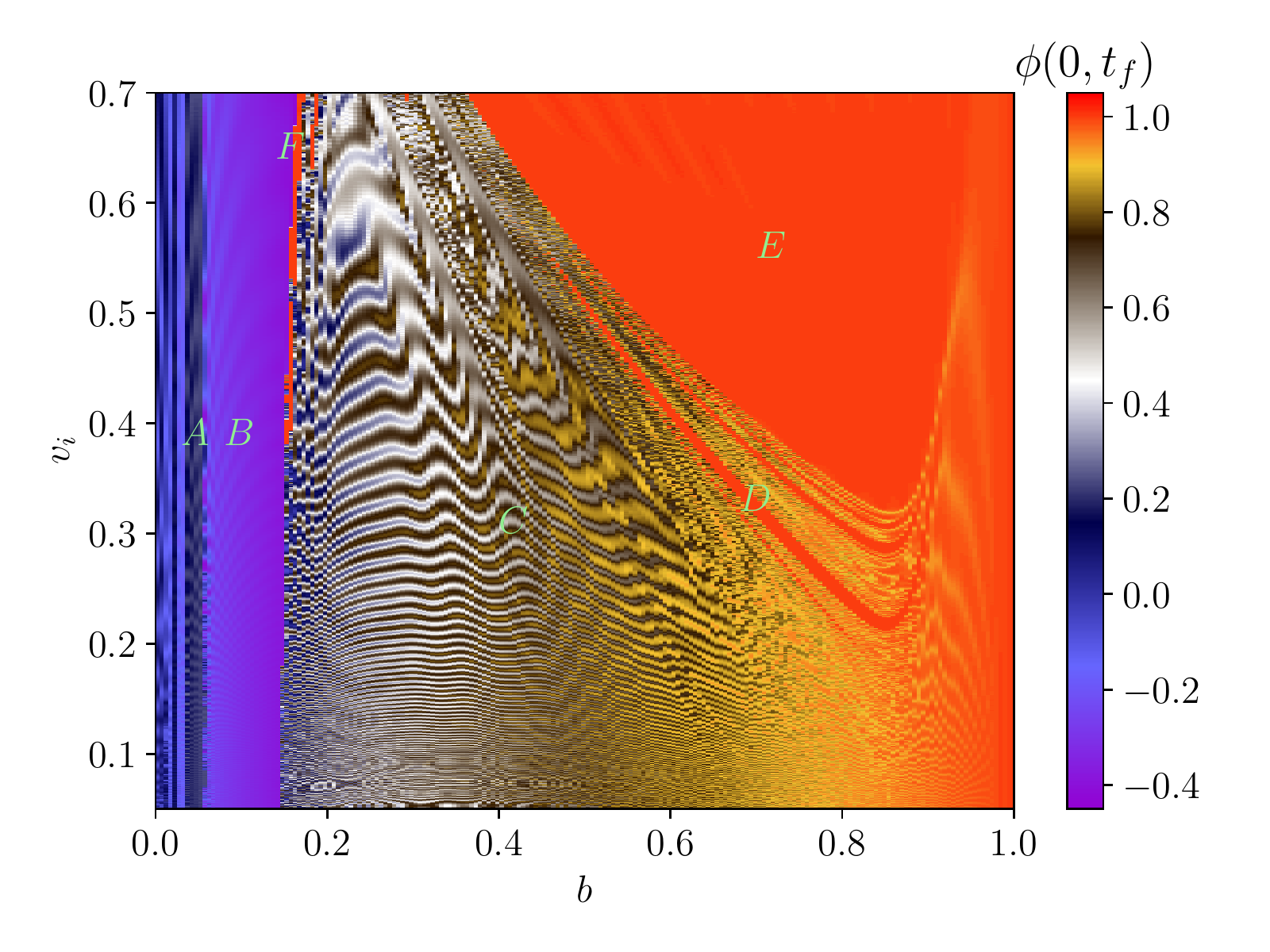}
    \caption{Final value of $\phi$ at the collision center as a function of $b$ and $v_i$. We are considering
kink-antikink collisions between outer kinks.}
    \label{fig:mat-out-ka}
\end{figure}

\begin{figure}
    \centering
    \begin{subfigure}{\textwidth} 
    \includegraphics[width=\textwidth]{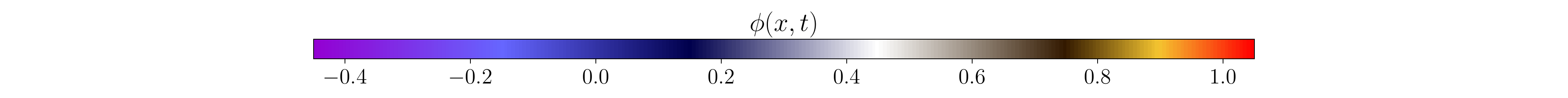}
    \end{subfigure}
    \begin{subfigure}{0.32\textwidth} 
    \caption{$b=0.045$, $v_i=0.38$.}
    \includegraphics[width=\textwidth]{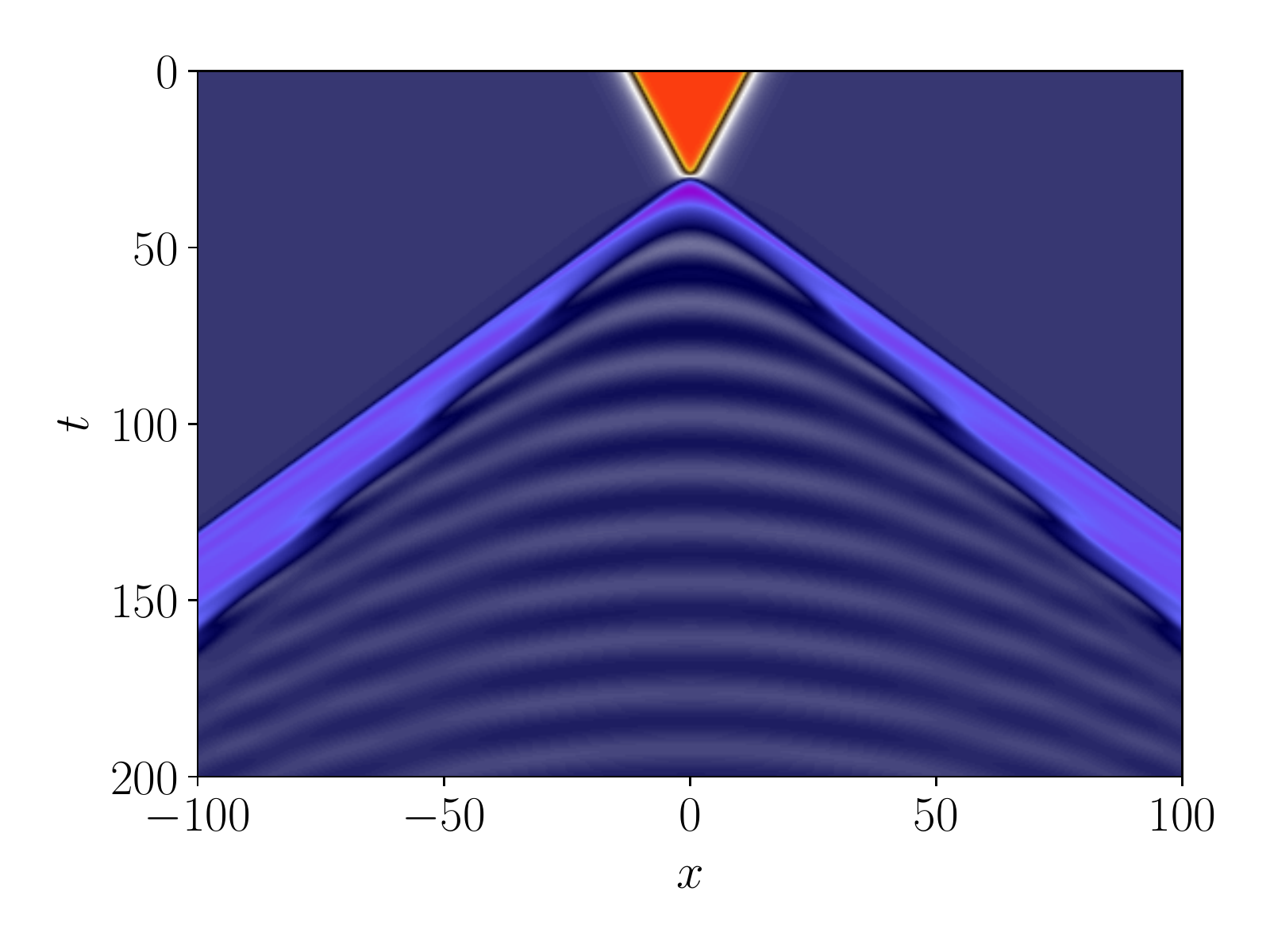}
    \end{subfigure}
    \begin{subfigure}{0.32\textwidth} 
    \caption{$b=0.08$, $v_i=0.38$.}
    \includegraphics[width=\textwidth]{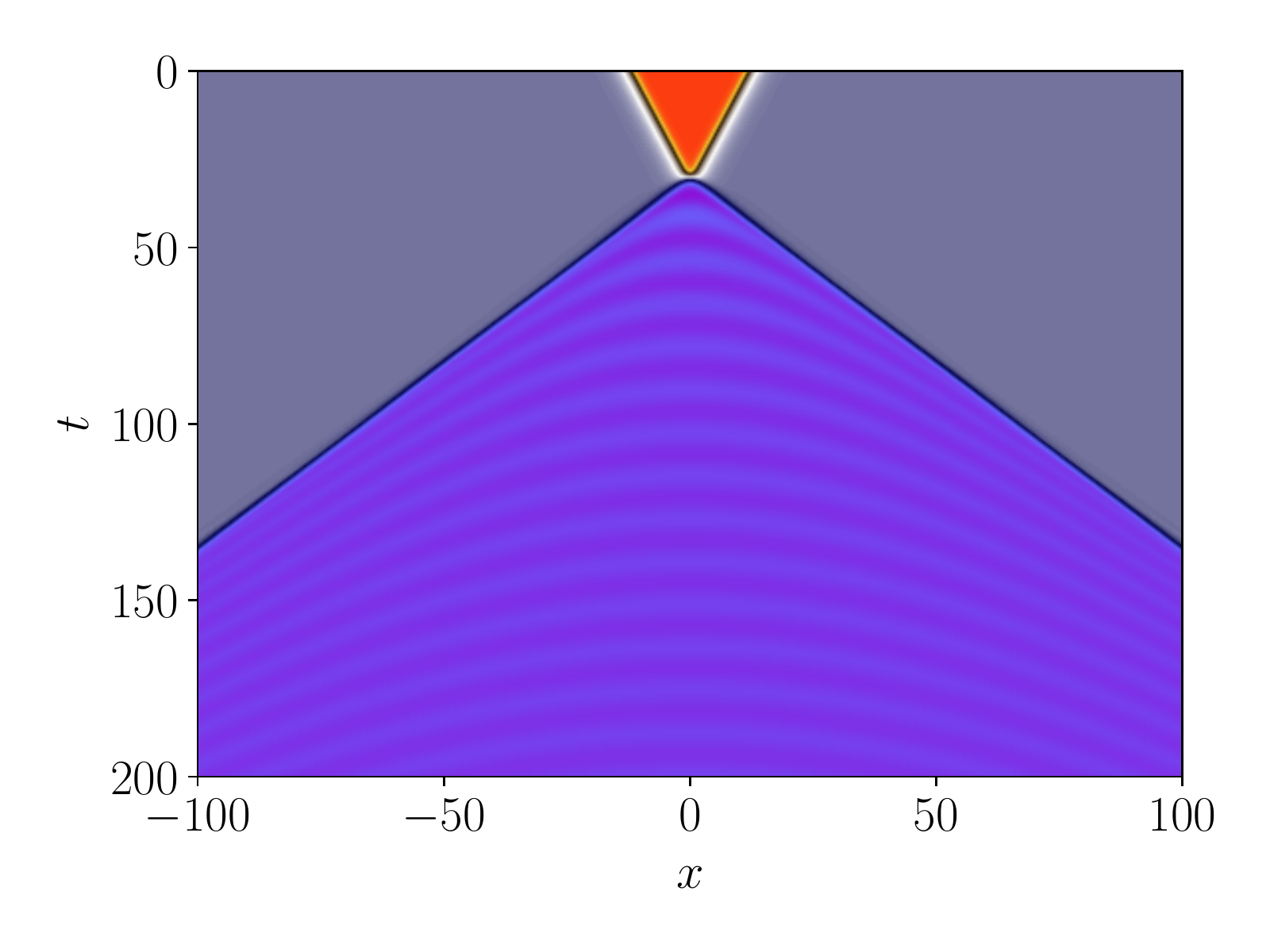}
    \end{subfigure}
    \begin{subfigure}{0.32\textwidth} 
    \caption{$b=0.4$, $v_i=0.3$.}
    \includegraphics[width=\textwidth]{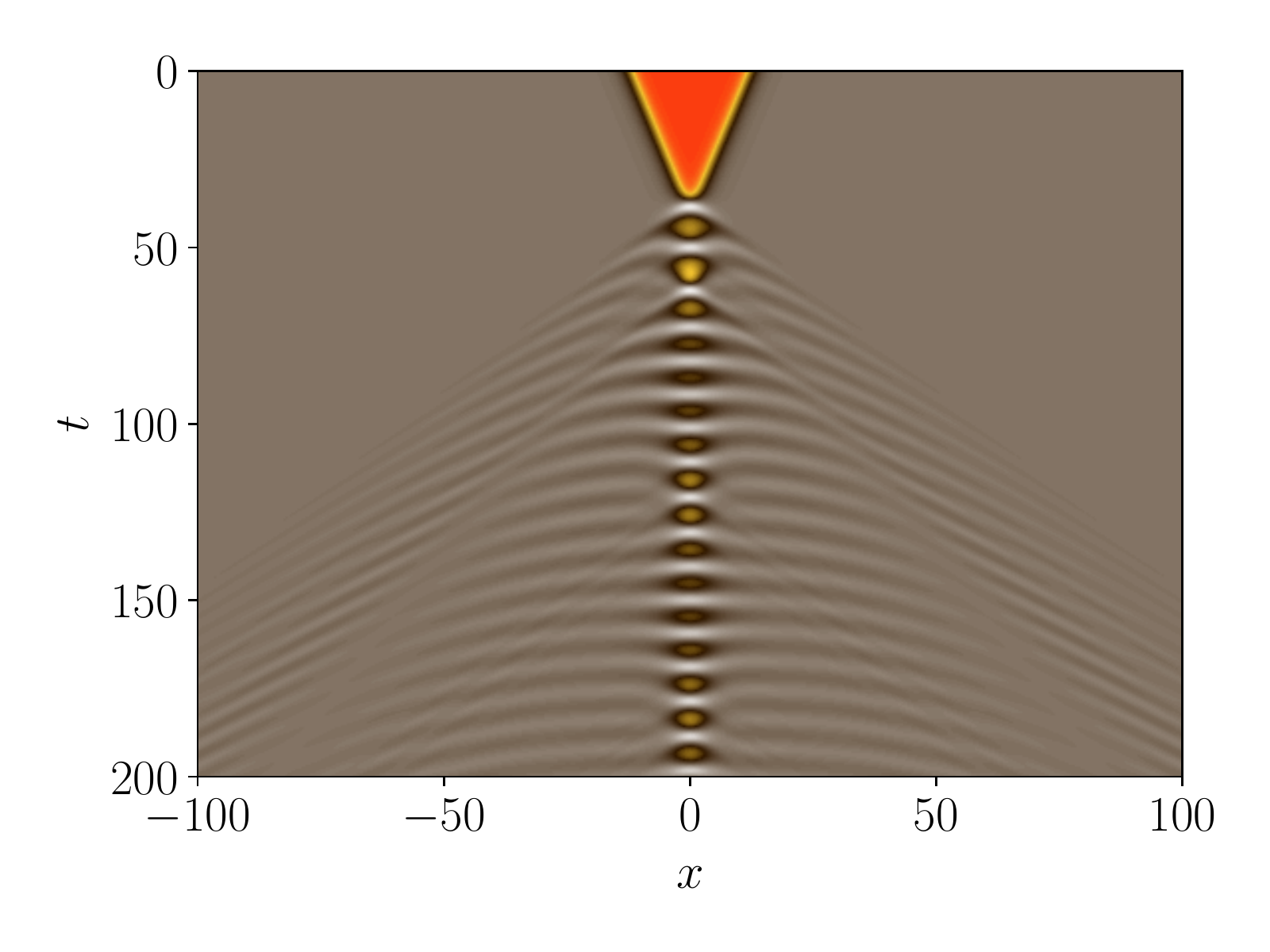}
    \end{subfigure}
    \begin{subfigure}{0.32\textwidth} 
    \caption{$b=0.67$, $v_i=0.34$.}
    \includegraphics[width=\textwidth]{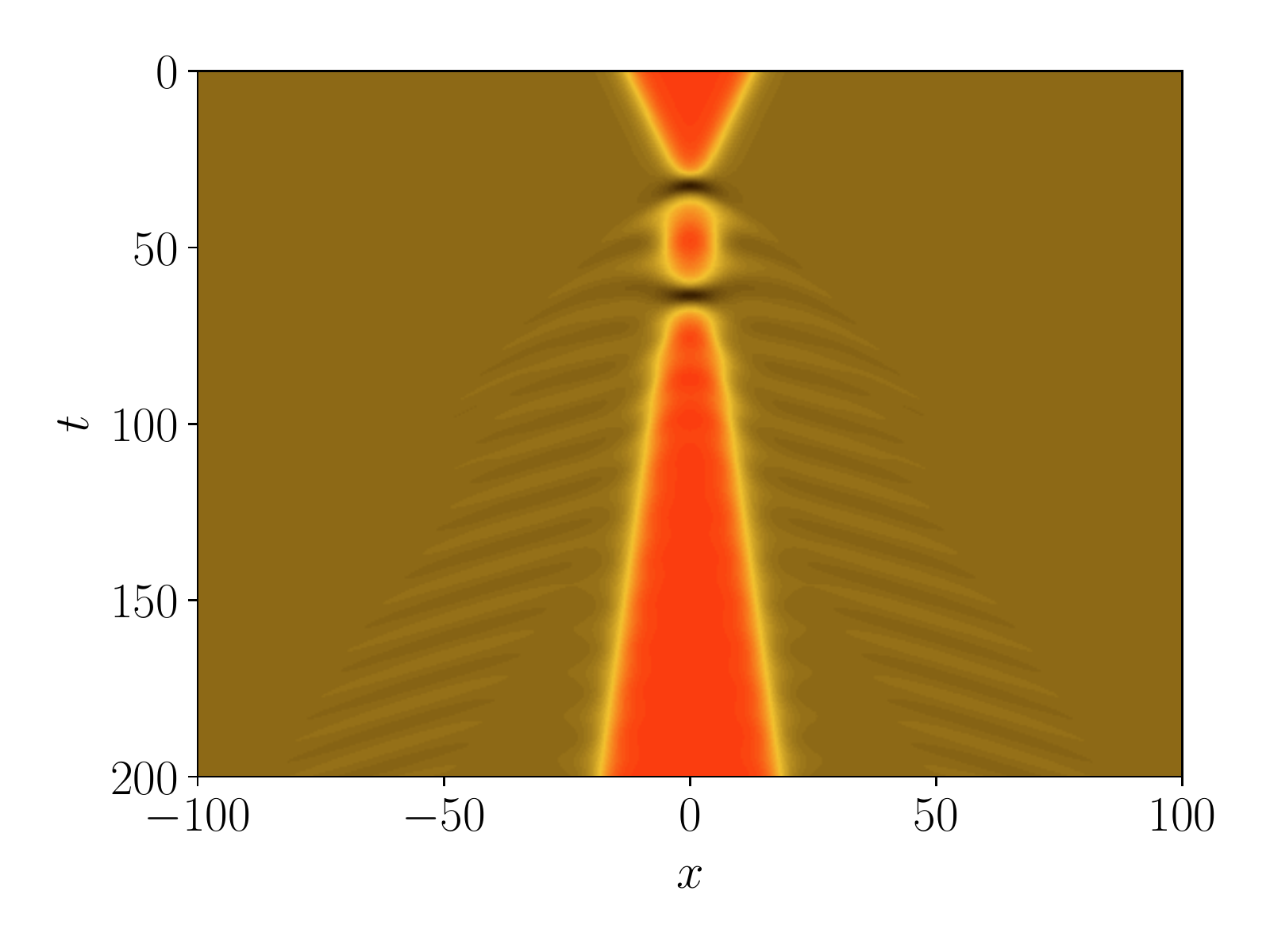}
    \end{subfigure}
    \begin{subfigure}{0.32\textwidth} 
    \caption{$b=0.7$, $v_i=0.55$.}
    \includegraphics[width=\textwidth]{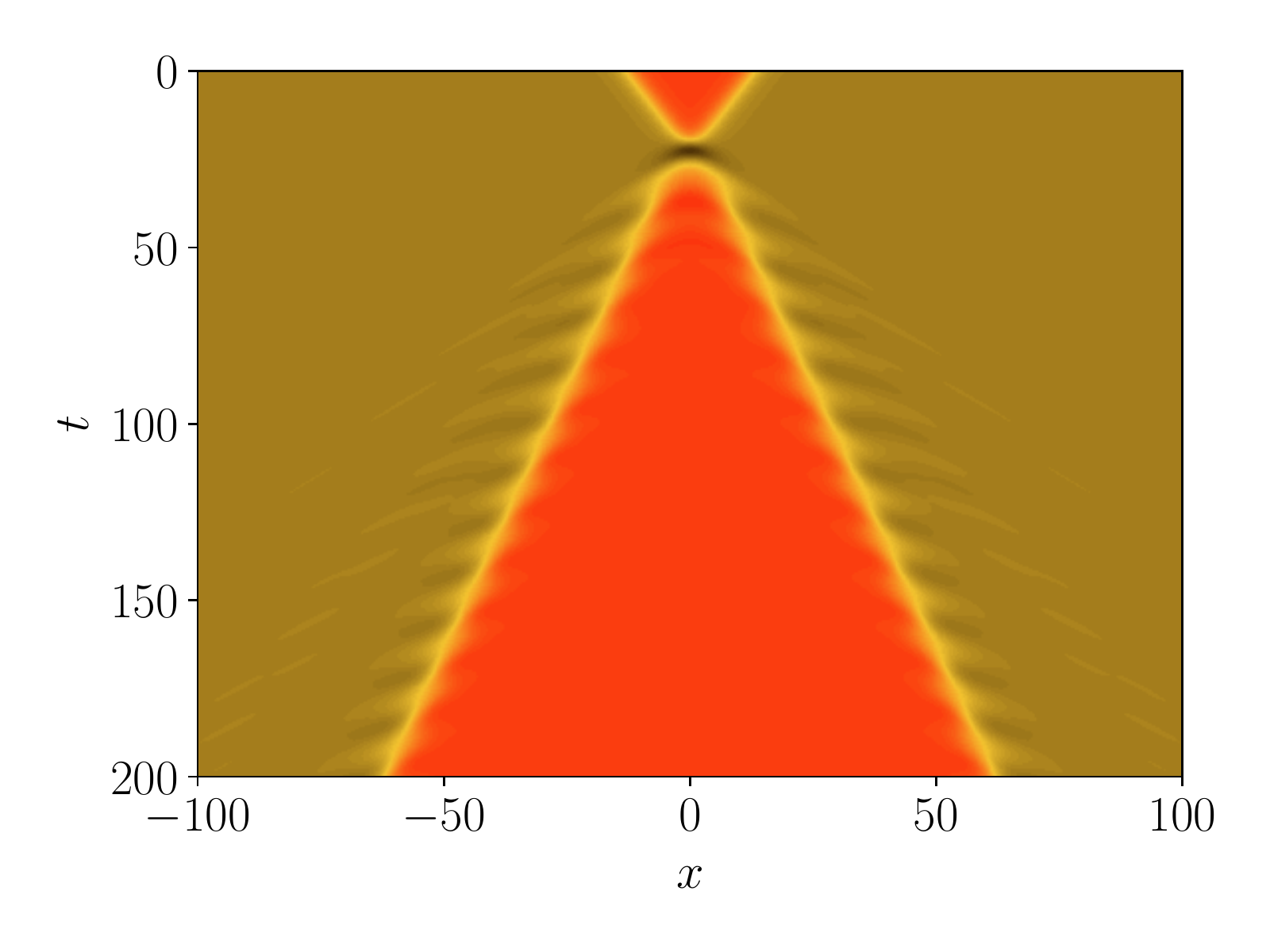}
    \end{subfigure}
    \begin{subfigure}{0.32\textwidth} 
    \caption{$b=0.16$, $v_i=0.64$.}
    \includegraphics[width=\textwidth]{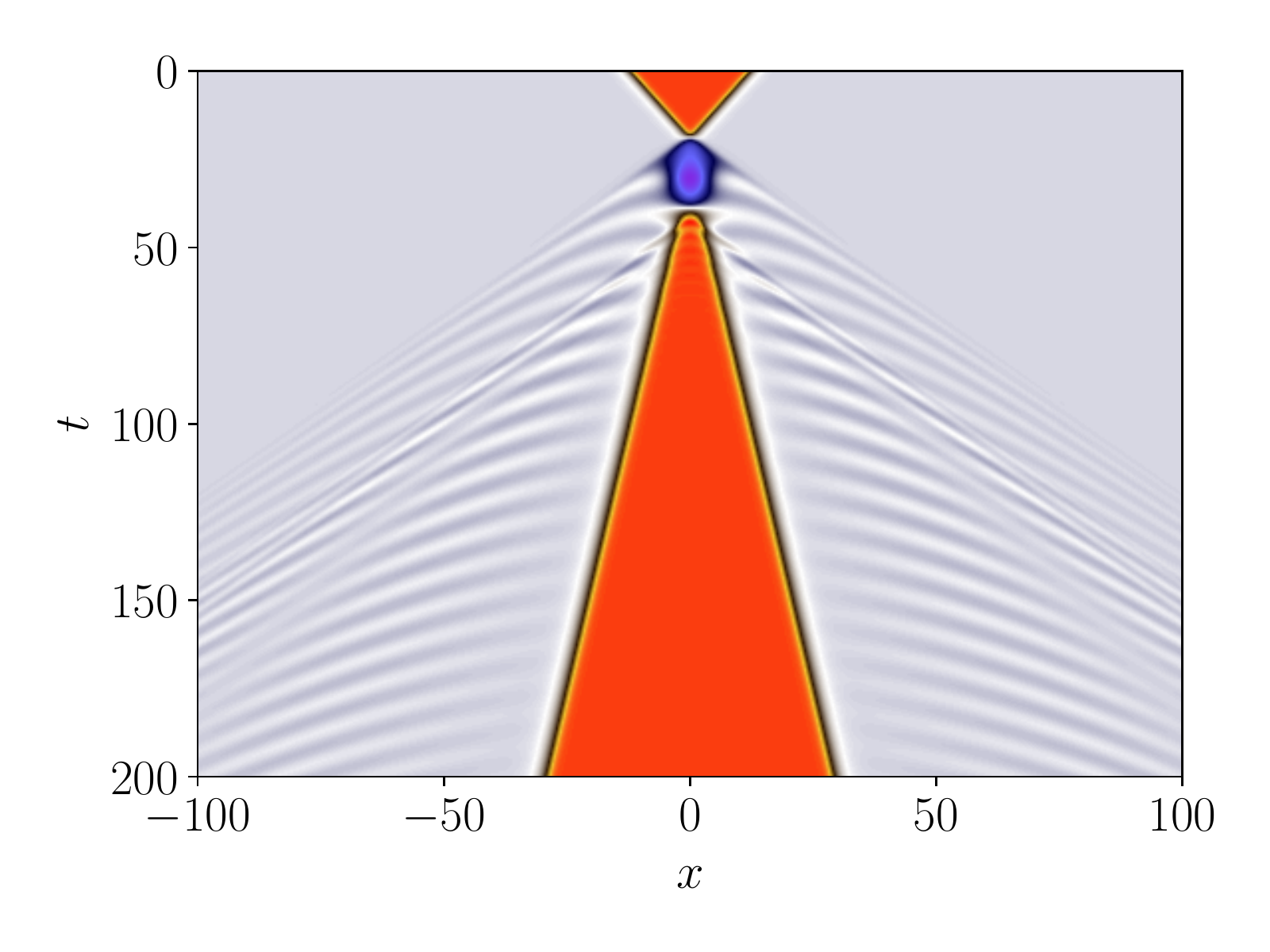}
    \end{subfigure}
    \caption{Evolution of $\phi(x,t)$ in spacetime for kink-antikink collisions between outer kinks.}
    \label{fig:field-out-ka}
\end{figure}

Fig.~\ref{fig:mat-out-ka} summarizes the system's behavior as a function of the parameter $b$ and the initial velocity $v_i$. It shows the field's value at the collision center at a time $t_f=60.0/v_i$. It is possible to distinguish six different regions, which will be listed below:
\begin{itemize}
    \item Region $A$: The kinks change sector, and an extra pair of inner kinks is created. It is illustrated in Fig.~\ref{fig:field-out-ka}(a).
    \item Region $B$: The kinks change sector after the collision and separate. It is illustrated in Fig.~\ref{fig:field-out-ka}(b).
    \item Region $C$: The kinks annihilate forming a bion as illustrated in Fig.~\ref{fig:field-out-ka}(c). 
    \item Region $D$: The kinks either form a bion or bounce more than once and then separate. The latter are resonance windows. The two-bounce case is illustrated in Fig.~\ref{fig:field-out-ka}(d).  
    \item Region $E$: The kinks collide and then reflect as illustrated in Fig.~\ref{fig:field-out-ka}(e).
    \item Region $F$: The kinks change sectors twice before separating as illustrated in Fig.~\ref{fig:field-out-ka}(f).
\end{itemize}

As $b\to 0$, we end up with an interesting situation. The tail that is not facing the opposite kink becomes long-range. As we move towards this limit, the formation of extra pairs of kinks from region $B$ to $A$ is completely analogous to the behavior of kink-antikink collisions in the inner sector. It occurs because the initial energy is sufficient to create more pairs as the inner kinks become lighter. Similarly, it eventually culminates in annihilation directly into radiation and an ultrarelativistic critical velocity. This is in agreement with the results obtained in Ref.~\cite{campos2021interaction}.


The inner kink is more massive in regions $D$ and $E$. Hence, it is not possible to change sectors anymore. Instead, we observe the standard resonance phenomenon, which is consistent with the presence of a vibrational mode. The region $F$ is at the boundary between the cases where it is possible to change sectors and where it is not. For the observed behavior to occur, the parameters need to be tuned. First, a sector change occurs, but the energy of the kinks in the new sector is very small. Thus, they do not separate and bounce back. Still, the energy is enough for the kinks to return to the original sector and separate. As a last observation, we see that the kinks become extremely small as $b$ approaches one, and we observe a sharp increase in the critical velocity again.

The simulations for $b\geq0.76$ were performed after minimizing the initial condition as described previously. Again, such a procedure was crucial to obtain correct results. In the limit $b\to0$, the tails become long-range, but the minimization procedure is unnecessary because they do not face the opposing kink.

\section{Conclusion}
\label{sec:conclusion}

We considered a class of $\phi^8$ scalar field models in (1+1)-dimensions. The scalar potentials which describe the theory contain four degenerate minima whose positions depend on one free parameter, $b$. In this scenario, a detailed analysis of kink-antikink collisions was made. This model exhibits two inner kinks related by parity or equivalently $\phi\to-\phi$ transformations. There are also four outer kinks related by parity and $\phi\to-\phi$ transformations. We computed the kink profiles and spectra numerically while the masses were computed analytically. Then, we analyzed kink-antikink collisions. Three types of kink-antikink collisions exist in the present model. We performed detailed numerical simulations of all cases and found different regions in parameter space, which were carefully analyzed. Here, we also used the minimization method proposed in Ref.~\cite{campos2021interaction} on a large scale, showing that it is highly computationally efficient.

We conclude that the $\phi^8$ model exhibits new phenomena with respect to lower-order polynomial models. It was already known that long-range tails occur. The case with four degenerate minima exhibits other phenomena and approaches the long-range case when $b=0$ and $b=1$. In particular, we found a rich fractal structure of resonance windows coming from both localized and delocalized modes. We also found sector change with the formation of oscillating pulses and extra kink-antikink pairs. The critical velocity exhibits a sharp increase in collisions between very small kinks. A fascinating case was obtained in Fig.~\ref{fig:field-out-ka}(f), where the kinks separate after changing sectors twice. More importantly, we provided a path from short-range to long-range behavior. When two quadratic minima merge, a large number of small kink-antikink pairs may form, creating a radiation pattern. This occurs when the tail not facing the opposing kink acquires long-range character. Therefore, our results confirm that, in such case, incoming kink and antikink form a large number of small kink-antikink pairs or decay directly into radiation.

One perspective for future works is analyzing in detail the rich structure of resonance windows that we found. For instance, it would be interesting to measure the resonant frequency of different sets of windows numerically and study their fractality. One could also study the $\phi^8$ model away from the first-order phase transition. As there are two parameters, one should be fixed to allow a similar analysis to the one performed here. We firmly believe that new and exciting phenomena will be encountered.
Furthermore, one could move on to the $\phi^{10}$ and $\phi^{12}$ models, which have three and four free parameters, respectively. The $\phi^{10}$ model with five degenerate minima has only one free parameter, and a similar analysis can be performed. However, two parameters need to be tuned to put the system in such a state. On the other hand, to fully comprehend the $\phi^{12}$ model is much more challenging. The case with six degenerate minima has two free parameters, and the remaining two must be tuned to put the system in such a state. Understanding these systems would be a fascinating challenge where new phenomena will be unveiled, and an even richer fractal structure may be found. It would also contribute to a better understanding of the value of having polynomial potentials with increasing power in the non-linearity of the field self-interaction.

\section*{Acknowledgments}

We acknowledge financial support from the National Council for Scientific and Technological Development - CNPq, Grant No. 303469/2019-6 (DB), No. 150166/2022-2 (JGFC), and No. 309368/2020-0 (AM). DB thanks Paraíba State Research Foundation, FAPESQ-PB, Grant N. 0015/2019. AM thanks financial support from the Brazilian agency CAPES and Universidade Federal de Pernambuco Edital Qualis A. The simulations presented here were conducted on the SDumont cluster at the Brazilian laboratory LNCC (Laborat\'orio Nacional de Computa\c{c}\~ao Cient\'ifica).

\end{document}